\documentclass[prd, amsfonts, twocolumn, nofootinbib, showpacs]{revtex4}
\usepackage{graphicx, epsfig}

\newcommand{\be}{\begin{equation}}
\newcommand{\ee}{\end{equation}}
\newcommand{\bea}{\begin{eqnarray}}
\newcommand{\eea}{\end{eqnarray}}

\newcommand\fdg{\hbox{$.\!\!^\circ$}}
\newcommand{\la}{\,\rlap{\raise 0.5ex\hbox{$<$}}{\lower 1.0ex\hbox{$\sim$}}\,}
\newcommand{\ga}{\,\rlap{\raise 0.5ex\hbox{$>$}}{\lower 1.0ex\hbox{$\sim$}}\,}

\def\etal{{\it et al.}}

\begin{document}

\title{
The Uncorrelated Universe: Statistical Anisotropy
and the Vanishing Angular Correlation Function in WMAP Years 1-3 
}

\author{Craig J. Copi$^1$}
\author{Dragan Huterer$^2$}
\author{Dominik J. Schwarz$^3$}
\author{Glenn D. Starkman$^{1,4}$}
\affiliation{$^1$ Department of Physics, Case Western Reserve University, 
             Cleveland, OH~~44106-7079}
\affiliation{$^2$ Kavli Institute for Cosmological Physics and Department of Astronomy 
  and Astrophysics, University of Chicago, Chicago, IL 60637}
\affiliation{$^3$ Fakult\"at f\"ur Physik, Universit\"at Bielefeld,
  Postfach 100131, 33501 Bielefeld, Germany}
\affiliation{$^4$ Beecroft Institute for Particle Astrophysics and Cosmology, Astrophysics, 
	University of Oxford, UK}

\begin{abstract}
  The large-angle (low-$\ell$) correlations of the Cosmic Microwave
  Background (CMB) as reported by the Wilkinson Microwave Anisotropy Probe
  (WMAP) after their first year of observations exhibited statistically
  significant anomalies compared to the predictions of the standard
  inflationary big-bang model.  
  We suggested then that these implied the presence of a solar system
  foreground, a systematic correlated with solar system geometry, or both.  We
  re-examine these anomalies for the data from the first three years of WMAP's
  operation.  We show that, despite the identification by the WMAP
  team of a systematic correlated with 
  the equinoxes and the ecliptic, 
  the anomalies in the first-year 
  Internal Linear Combination (ILC) map persist in the three-year ILC map,  
  in all-but-one case at similar statistical significance.
  The three-year ILC quadrupole and octopole therefore remain 
  inconsistent with statistical isotropy -- they are
  correlated with each other ($99.6\%$C.L.), and there are statistically
  significant correlations with local geometry, 
  especially that of the solar system.  
  The angular two-point correlation function at scales $>60$
  degrees in the regions outside the (kp0) galactic cut, where it is most
  reliably determined, is approximately zero in all wavebands and is even
  more discrepant with the best fit $\Lambda$CDM inflationary model than 
  in the first-year data -- $99.97\%$C.L.~for the new ILC map.  
  The full-sky ILC map, on the other hand, has a
  non-vanishing angular two-point correlation function, apparently driven by
  the region inside the cut, but which does not agree better with $\Lambda$CDM.
  The role of the newly identified low-$\ell$ systematics is more puzzling 
  than reassuring. 
\end{abstract}


\maketitle

\section{Introduction and results}
Approximately three years ago, the Wilkinson Microwave Anisotropy Probe 
(WMAP) team reported 
\cite{WMAP1_results,WMAP1_foreground,WMAP1_angps,WMAP1_TE,WMAP1_low} 
the results of its preliminary analysis of the satellite's first year 
of operation. While the data is regarded as a dramatic confirmation of 
standard inflationary cosmology, anomalies exist.  

Among the unexpected first-year results, there is a natural division 
into several classes:   
(1) a lack of large angle correlations \cite{WMAP1_results}, 
and violations of statistical isotropy in the associated multipoles --- 
$\ell=2$ and $3$ \cite{deOliveira-Costa:2003pu, SSHC}; 
plus weaker evidence for a violation of statistical isotropy at $\ell =6$ and 
$7$~\cite{Freeman:2005nx};   
(2) statistically anomalous values for the angular power spectrum, 
$C_\ell$ in at least three $\ell$ bins:
a trough at $\ell=22(20-24)$, a peak at $\ell=40(37-44)$, 
and a trough at $\ell=210(201-220)$~\cite{WMAP1_angps};
(3) hemispheric asymmetries in the angular power spectrum over a wide
range of $\ell$ \cite{NS_asymmetry};
(4) unexpectedly high cross-correlation between temperature and 
E-mode polarization (TE) at low $\ell$ \cite{WMAP1_results},  
interpreted by the WMAP team as evidence for a large optical depth 
and thus for very early star formation \cite{WMAP1_low}; and 
a discrepancy between the observed TE angular power at the largest 
scales and the best fit concordance model \cite{Dore03}. 

In this paper, we focus on the large angle anomalies of the microwave
background as measured in the satellite's first three years of operation,
and recently reported by the WMAP team.
In particular, we look at how the full three-year
WMAP results \cite{WMAP123_results,WMAP123_beam,
WMAP123_angps,WMAP123_pol} (hereafter WMAP123) 
are similar to or different from those
obtained from the analysis of first-year data (WMAP1). 
Given the high signal-to-noise of the first-year WMAP maps
over a large number of  well distributed pixels, 
in the absence of newly-identified systematic effects,
one would not expect significant changes in the low-$\ell$ multipoles 
between the one-year and three-year data.  
The recent detailed analysis of \cite{deOliveira-Costa06} quantified this expectation,
but suggested that the quadrupole, being anomalously low, 
is the least robust of the low multipoles. 

In Section II we repeat our previous analysis of the CMB data using the
multipole vector framework \cite{SSHC,CHSS}.  In Section III we recall the
lack of angular correlation at large angles, as measured by COBE-DMR
\cite{DMR4} and confirmed by WMAP1 \cite{WMAP1_low}, but not investigated
in the recent analysis of the WMAP team.  We therefore include our own
preliminary analysis of the angular two-point correlation function in this
paper. We discuss the consequences of our findings in Section IV.

For the multipole vector analysis we compute the quadrupole and octopole
multipole vectors and associated area vectors \cite{vectors} of the WMAP123
full sky map (ILC123) \cite{WMAP123_products}, and compare them with those
derived \cite{SSHC,CHSS} from three full sky maps based on WMAP1 data ---
WMAP's own ILC1 map \cite{WMAP1_products}; Tegmark, de Oliveira-Costa and
Hamilton's (TOH1) map \cite{Tegmark:2003ve}; and the Lagrange multiplier
Internal Linear Combinations map (LILC1) \cite{LILC1}.  We look at the
correlations among the quadrupole and octopole area vectors, and confirm
that in the ILC123 these are anomalously aligned ($\ga 99.5\%$C.L.), as
found \cite{SSHC,CHSS} in the ILC1, TOH1 and LILC1 ($99.4\%$C.L.;
confirming and strengthening earlier results in
\cite{deOliveira-Costa:2003pu}).  We confirm that, as for the first-year
maps, these alignments are strengthened by proper removal of the Doppler
contribution to the quadrupole \cite{SSHC}.

Having established the mutual alignment of the four planes defined by the
quadrupole and octopole, we recall that in the WMAP1 maps it was found
\cite{SSHC,CHSS} that these planes were aligned not just among themselves, but
also with several physical directions or planes.  As in the previous work we
study these alignments both with and without accepting the internal
correlations between the quadrupole and octopole as given.  We find that the
alignments generally retain the same significance as in the first-year all-sky
maps.  Only the alignment between the ecliptic plane and the quadrupole and
octopole planes, especially given the latter's internal correlations, is
noticeably weakened.  Meanwhile, there persist further correlations of the
ecliptic and the quadrupole-plus-octopole, which rely strongly on the alignment
of the quadrupole and octopole planes with the ecliptic.

Finally we also examine the map of the combined quadrupole and
octopole and observe that, as for WMAP1, the ecliptic plane traces a zero
of the map over approximately one-third of the sky, and separates the two
strongest extrema in the southern ecliptic hemisphere from the two weakest
extrema in the northern hemisphere. 

Given the evidence for the lack of statistical isotropy on large scales we
provide a careful description of the definition of the angular two-point
correlation function means and connections between different possible
definitions of this function.  Our analysis shows that the angular two-point
correlation function is strikingly deficient at large angles -- in fact, the
deficit of power is even more significant in the new data than in the first
year data ($99.97\%$C.L.~for the ILC123 in the region outside the kp0 galaxy
mask). Moreover, we find that the quadrupole and octopole power computed
directly from the cut-sky maps are inconsistent with the quadrupole and
octopole computed from the maximum likelihood estimators (the latter then being
used for all cosmological analyses).  In the conclusions section we discuss the
implications of these findings.

\section{Multipole Vectors}
\begin{table*}
  \caption{
    Multipole vectors, $\hat v^{(\ell,i)}$, and area vectors ${\vec
      w}^{(\ell;i,j)}$, for the quadrupole and octopole in Galactic
    coordinates $(l,b)$.  (Magnitudes are also given for the area vectors.)
    All vectors are given for the ILC123, ILC1, TOH1, and LILC1 maps after
    correcting for the kinetic quadrupole.
  }
  \label{tab:ILC123vectors}
\begin{tabular}{lr@{$\fdg$}lr@{$\fdg$}lc@{\hspace{3em}}lr@{$\fdg$}lr@{$\fdg$}lc}
\hline
\multicolumn{1}{c}{Vector} & \multicolumn{2}{c}{$l$} &
\multicolumn{2}{c}{$b$} & Magnitude &
\multicolumn{1}{c}{Vector} & \multicolumn{2}{c}{$l$} &
\multicolumn{2}{c}{$b$} & Magnitude \\ \hline
\multicolumn{6}{c}{ILC123} & \multicolumn{6}{c}{ILC1} \\ \hline
%
$\hat v^{(2,1)}$ & $119$&$6$ & $10$&$8$ & --- &
$\hat v^{(2,1)}$ & $115$&$2$ & $23$&$6$ & --- \\
%
$\hat v^{(2,2)}$ & $5$&$9$ & $19$&$6$ & --- &
$\hat v^{(2,2)}$ & $19$&$5$ & $8$&$6$ & --- \\
$\vec w^{(2;1,2)}$ & $-128$&$3$ & $63$&$0$ & 0.951 &
$\vec w^{(2;1,2)}$ & $-88$&$9$ & $64$&$4$ & 0.999 \\
%
$\hat v^{(3,1)}$ & $93$&$8$ & $39$&$5$ & --- &
$\hat v^{(3,1)}$ & $95$&$3$ & $37$&$0$ & --- \\
$\hat v^{(3,2)}$ & $23$&$9$ & $8$&$3$ & --- &
$\hat v^{(3,2)}$ & $21$&$7$ & $9$&$4$ & --- \\
$\hat v^{(3,3)}$ & $-46$&$3$ & $11$&$7$ & --- &
$\hat v^{(3,3)}$ & $-47$&$0$ & $10$&$7$ & --- \\
$\vec w^{(3;1,2)}$ & $-76$&$1$ & $50$&$0$ & 0.936 &
$\vec w^{(3;1,2)}$ & $-80$&$9$ & $52$&$9$ & 0.947 \\
$\vec w^{(3;2,3)}$ & $154$&$9$ & $77$&$5$ & 0.934 &
$\vec w^{(3;2,3)}$ & $161$&$7$ & $77$&$8$ & 0.924 \\
$\vec w^{(3;3,1)}$ & $-143$&$9$ & $32$&$9$ & 0.892 &
$\vec w^{(3;3,1)}$ & $-144$&$3$ & $33$&$9$ & 0.861 \\
\hline
\multicolumn{6}{c}{TOH1} & \multicolumn{6}{c}{LILC1} \\ \hline
%
$\hat v^{(2,1)}$ & $118$&$9$ & $25$&$1$ & --- &
$\hat v^{(2,1)}$ & $125$&$8$ & $16$&$4$ & --- \\
$\hat v^{(2,2)}$ & $11$&$2$ & $16$&$6$ & --- &
$\hat v^{(2,2)}$ & $5$&$9$ & $17$&$6$ & --- \\
$\vec w^{(2;1,2)}$ & $-105$&$7$ & $56$&$6$ & 0.990 &
$\vec w^{(2;1,2)}$ & $-115$&$3$ & $58$&$5$ & 0.929 \\
%
$\hat v^{(3,1)}$ & $86$&$9$ & $39$&$3$ & --- &
$\hat v^{(3,1)}$ & $89$&$2$ & $37$&$7$ & --- \\
$\hat v^{(3,2)}$ & $22$&$6$ & $9$&$2$ & --- &
$\hat v^{(3,2)}$ & $23$&$8$ & $9$&$7$ & --- \\
$\hat v^{(3,3)}$ & $-44$&$9$ & $8$&$2$ & --- &
$\hat v^{(3,3)}$ & $-47$&$3$ & $10$&$6$ & --- \\
$\vec w^{(3;1,2)}$ & $-78.4$&$6$ & $49$&$8$ & 0.902 &
$\vec w^{(3;1,2)}$ & $-78$&$6$ & $51$&$7$ & 0.904 \\
$\vec w^{(3;2,3)}$ & $173$&$8$ & $79$&$5$ & 0.918 &
$\vec w^{(3;2,3)}$ & $164$&$5$ & $77$&$6$ & 0.939 \\
$\vec w^{(3;3,1)}$ & $-141$&$6$ & $38$&$9$ & 0.907 &
$\vec w^{(3;3,1)}$ & $-145$&$4$ & $36$&$8$ & 0.892 \\
\hline
\end{tabular}
\end{table*}

Several large-angle anomalies or putative anomalies in the CMB data have
been pointed out and discussed extensively in the literature
\cite{deOliveira-Costa:2003pu, NS_asymmetry, vectors, Chiang03, Park03,
  Dore03, Vielva03, Jain03, Gurzadyan03, SSHC, McEwen,
  Mukherjee04, Donoghue, Bielewicz04, LandMagueijo04a, LandMagueijoAoE, LandMagueijo05c, 
  CHSS, Bernui}.
Similarly, several novel methods proposed to study the statistical isotropy
or Gaussianity of the CMB have also been introduced and discussed in the
literature \cite{Hajian,ArmendarizPicon,MedeirosContaldi,Chiang06}.  Here
we focus on our previous studies \cite{SSHC, CHSS}, namely the alignment of
the quadrupole and octopole with each other and with physical directions on
the sky as revealed by the multipole vector formalism \cite{vectors}.

The $\ell$-th multipole of the CMB,
$T_\ell$, can, instead of being expanded in spherical harmonics,
be written uniquely \cite{Maxwell,vectors,Dennis2004} 
in terms of a scalar $A^{(\ell)}$
which depends only on the total power in this multipole and $\ell$ unit
vectors $\{ {\hat v}^{(\ell,i)}\vert i=1,...,\ell\}$.  These ``multipole
vectors'' encode all the information about the phase relationships of the
$a_{\ell m}$.  Heuristically,
\begin{equation}
  T_{\ell} \approx 
  A^{(\ell)} \prod_{i=1}^{\ell}({\hat v}^{(\ell,i)}\cdot {\hat e}) \, ,
\end{equation}
where ${\hat v}^{(\ell,i)}$ is the $i^{\rm th}$ multipole vector of the 
$\ell^{\rm th}$ multipole. (In fact the right hand side contains terms with 
``angular momentum'' $\ell-2$, $\ell-4$, ... These are subtracted by taking 
the appropriate traceless symmetric combination, as described in 
\cite{vectors, SSHC} and \cite{CHSS}.) 
Note that the signs of all the vectors can be absorbed into the sign of $A^{(\ell)}$,
so one is free to choose the hemisphere of each vector.  Unless otherwise noted,
we will choose the north galactic hemisphere when quoting the co-ordinates
of the multipole vectors, but in plots we will show the vector in both
hemispheres.

\subsection{Multipole  Vectors: WMAP1 vs.~WMAP123}
Multipole vectors are best calculated on cleaned full sky maps.  
The ILC1, TOH1, and LILC1 are all minimum-variance maps obtained from
WMAP's single-frequency maps, but differ in the detailed implementation of this
idea. These full-sky maps may have residual foreground contamination, probably
mainly due to imperfect subtraction of the Galactic signal.  They also have
complicated noise properties \citep{WMAP1_foreground} that make them less than
ideal for cosmological tests.  While one can, in principle, straightforwardly
compute the true (full-sky) multipole vectors from the single-frequency maps
with a sky cut, a cut larger than a few degrees across will introduce
significant uncertainty in the reconstructed full-sky multipole vectors and
consequently in any statistics that use them (see Section 7 and Figure 7 of
\cite{CHSS}).

\begin{figure*}
	\includegraphics[width=4in,angle=-90]{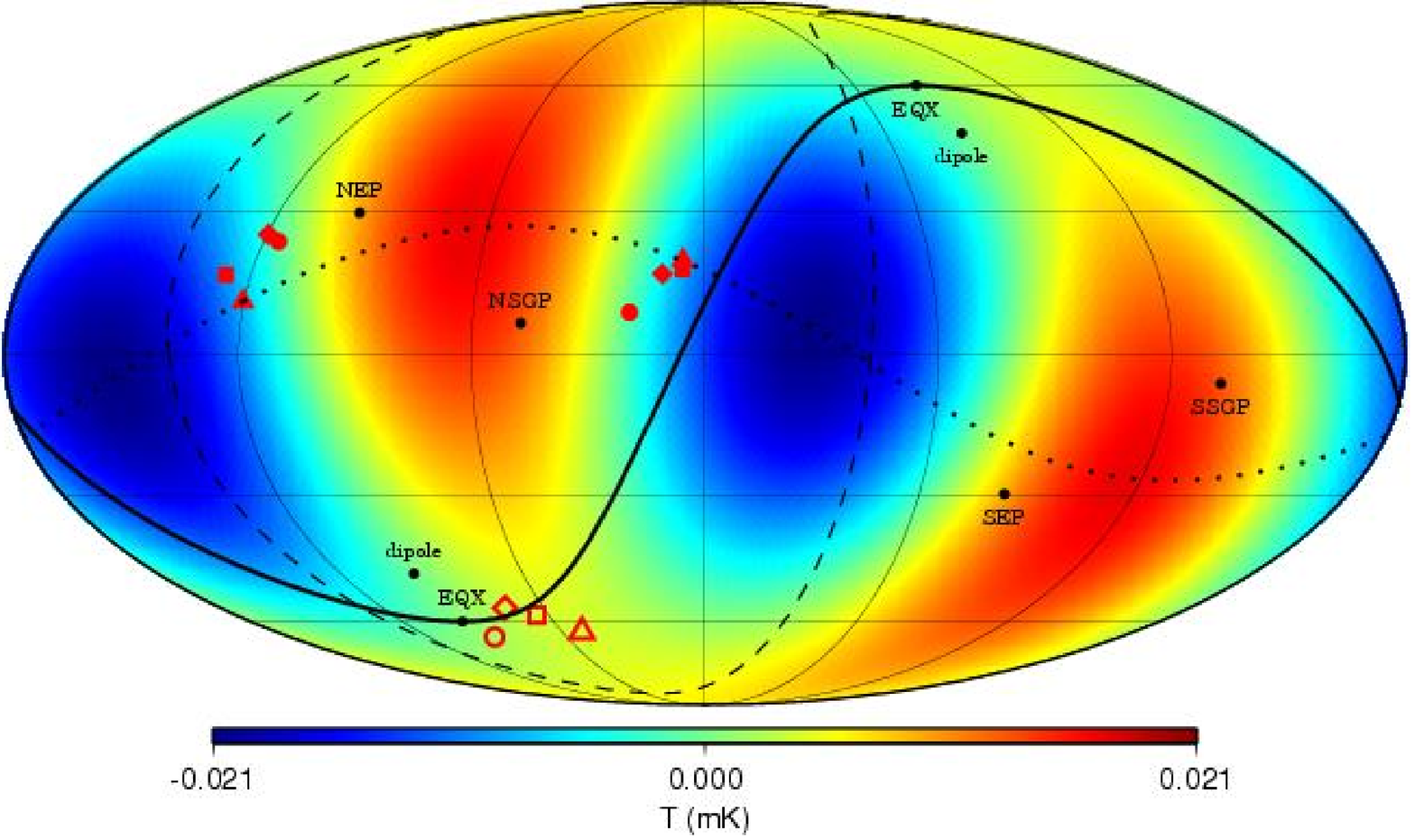}
	\includegraphics[width=4in,angle=-90]{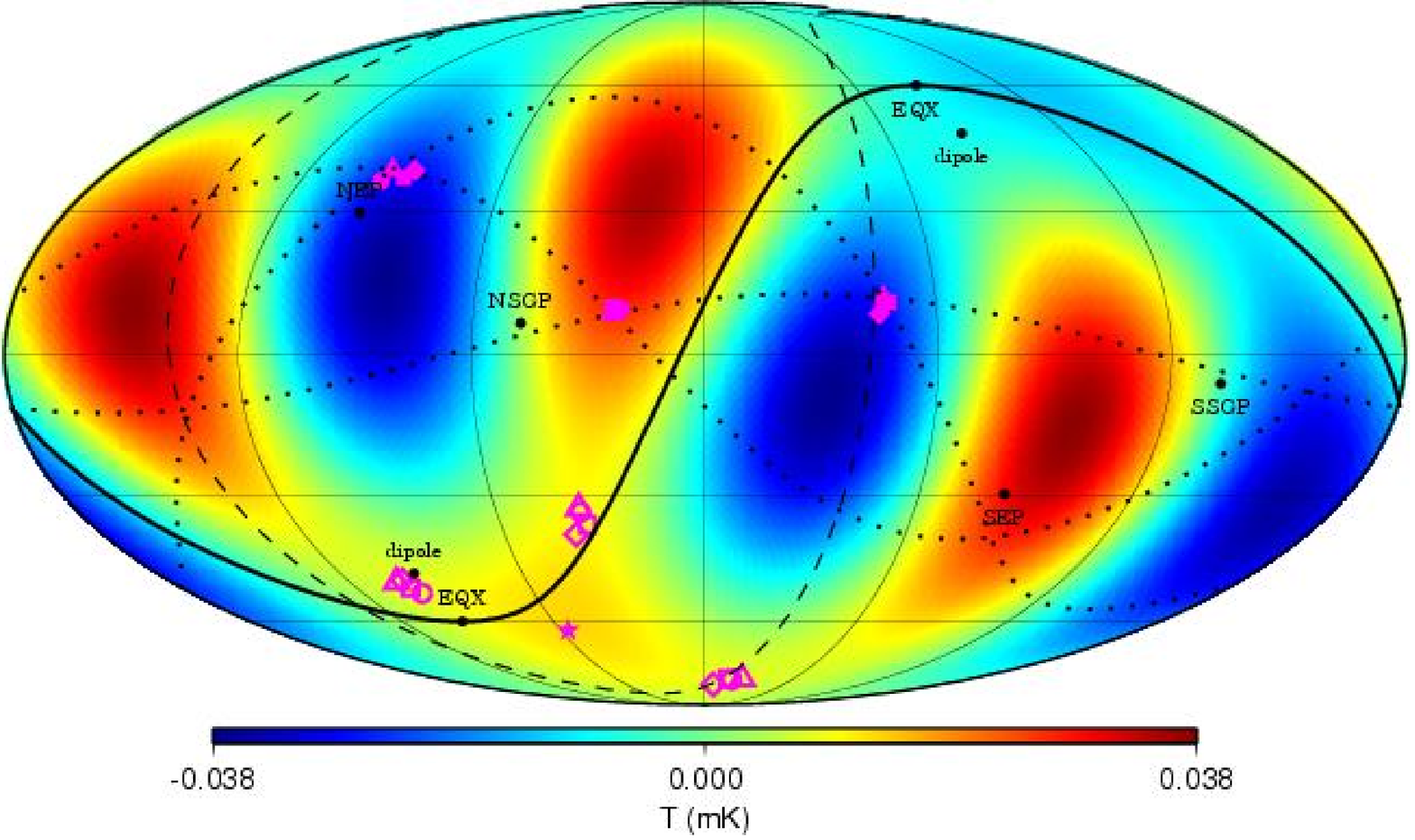}
	\caption{The $\ell=2$ (top panel) and $\ell=3$ (bottom panel)
	multipoles from the ILC123  cleaned map, presented in Galactic coordinates, 
	after correcting for the kinetic quadrupole.  The
	solid line is the ecliptic plane and the dashed line is the supergalactic
	plane.  The directions of the equinoxes (EQX), dipole due to our motion
	through the Universe, north and south ecliptic poles (NEP and SEP) and
	north and south supergalactic poles (NSGP and SSGP) are shown.  The
	multipole vectors are plotted as the solid red  symbols for $\ell=2$ 
	and solid magenta for $\ell=3$ (dark and medium gray in gray scale versions)
	for each map, ILC1 (circles), ILC123 (triangles), TOH1 (diamonds)
        and LILC1 (squares).  
	The open symbols of the same shapes are for the normal vectors
	for each map.  The dotted lines are the great circles connecting each pair of 
	multipole vectors for the ILC123 map.  
	For $\ell=3$ (bottom panel), the solid magenta (again medium gray
        in the gray scale version) star is the direction of the maximum
        angular momentum dispersion axis for the ILC123 octopole.
      }
	\label{fig:map:ilc123:2and3}
\end{figure*}

The multipole vectors for $\ell=2$ and $3$ 
for the Doppler-quadrupole-corrected ILC1, TOH1, LILC1 and ILC123 maps 
are given in Table \ref{tab:ILC123vectors} and plotted in
Figure~\ref{fig:map:ilc123:2and3}.
It is important that the Doppler contribution to the quadrupole, 
inferred from the measured dipole, has been removed before computing the multipole vectors.
Although the  Doppler-induced piece of the quadrupole is a small part of the total power,
as shown in \cite{SSHC} and \cite{CHSS}, failure to properly correct the quadrupole
for the Doppler contribution results in reduced significance for the quadrupole-octopole
correlations in all full sky maps.
Freeman \etal~\cite{Freeman:2005nx} have suggested that both
the lack of low-$\ell$ power and certain reported violations of statistical 
isotropy (in particular any north-south asymmetry) reported in cut-sky
analysis could be due to the use of a wrong value for the dipole. 
However, uncertainty in the dipole seems not to significantly affect the 
correlations in quadrupole and octopole multipole vectors discussed here,
as in the full-sky maps the identification of residual dipoles can be
done uniquely.

\begin{table} 
\caption{ Comparisons between ILC1 and ILC123 multipole vectors and area vectors.
Three quantities are tabulated:
(a) dot products $d^{(\ell,i)}$ between the multipole vectors from the ILC1,
${\hat v}_{\rm ILC1}^{(\ell,i)}$ and the corresponding vectors from the ILC123,
${\hat v}_{\rm ILC123}^{(\ell,i)}$, for $\ell=2$ through $\ell=7$;
(b) dot products $\Delta^{(\ell;i,j)}$ between the normalized area vectors from the ILC1,
${\hat w}_{\rm ILC1}^{(\ell;i,j)}$ and the corresponding area vectors from the ILC123,
${\hat w}_{\rm ILC123}^{(\ell;i,j)}$, for $\ell=2$ and $\ell=3$; 
(c) ratios $r^{(\ell;i,j)} $ between the magnitude of the area vectors from the ILC1
$\vert{\vec w}_{\rm ILC1}^{(\ell;i,j)}\vert$ and the corresponding magnitudes
$\vert{\vec w}_{\rm ILC123}^{(\ell;i,j)}\vert$ from the ILC123,for $\ell=2$ and $\ell=3$. 
For comparison, we show in column 3  the same quantity for ILC1 versus TOH1.
}
\label{tab:ILC123dotproducts}
\begin{tabular}{lcc}
\
Vector \quad\quad  &\quad  ILC1-ILC123 \quad & \quad ILC1-TOH1\quad \\ 
\hline
$d^{(2,1)}$ 	& 0.973	& 0.998	\\
$d^{(2,2)}$ 	& 0.956	& 0.980	\\
$\Delta^{(2;1,2)}$ &0.955 & 0.981	\\
$r^{(2;1,2)}$ 	& 0.951	& 1.010	\\
\hline
$d^{(3,1)}$ 	& 0.999	& 0.993	\\
$d^{(3,2)}$ 	& 0.999	& 1.000	\\
$d^{(3,3)}$ 	& 1.000	& 0.998	\\
$\Delta^{(3;1,2)}$ & 0.997 & 0.998	\\
$\Delta^{(3;2,3)}$ & 1.000 & 0.999	\\
$\Delta^{(3;3,1)}$ & 1.000 & 0.995	\\
$r^{(3;1,2)}$ 	& 0.988	& 1.050	\\
$r^{(3;2,3)}$ 	& 1.010	& 1.006	\\
$r^{(3;3,1)}$ 	& 1.036	& 0.949	\\
\hline
$d^{(4,1)}$ 	& 0.999	& 0.998	\\
$d^{(4,2)}$ 	& 0.996	& 0.981	\\
$d^{(4,3)}$ 	& 0.988	& 0.998	\\
$d^{(4,4)}$ 	& 0.998	& 0.993	\\
\hline
$d^{(5,1)}$ 	& 0.999	& 0.999	\\
$d^{(5,2)}$ 	& 0.999	& 1.000	\\
$d^{(5,3)}$ 	& 1.000	& 0.998	\\
$d^{(5,4)}$ 	& 1.000	& 0.997	\\
$d^{(5,5)}$ 	& 1.000	& 0.996	\\
\hline
$d^{(6,1)}$ 	& 0.995	& 0.979	\\
$d^{(6,2)}$ 	& 0.983	& 0.993	\\
$d^{(6,3)}$ 	& 0.999	& 0.999	\\
$d^{(6,4)}$ 	& 0.997	& 0.983	\\
$d^{(6,5)}$ 	& 0.993 & 0.987	\\
$d^{(6,6)}$ 	& 1.000	& 0.999	\\
\hline
$d^{(7,1)}$ 	& 0.996 & 0.997	\\
$d^{(7,2)}$ 	& 0.998	& 0.994	\\
$d^{(7,3)}$ 	& 1.000	& 0.998	\\
$d^{(7,4)}$ 	& 0.998	& 0.979	\\
$d^{(7,5)}$ 	& 1.000	& 1.000	\\
$d^{(7,6)}$ 	& 0.999	& 0.997	\\
$d^{(7,7)}$ 	& 1.000	& 0.999	\\
\hline
\end{tabular} 
\end{table} 

We have found that the area vectors 
\be
{\vec w}^{(\ell;i,j)}\equiv {\hat v}^{(\ell,i)} \times {\hat v}^{(\ell,j)}  .
\ee
are often more useful than the multipole vectors themselves for statistical
comparison. As discussed in \cite{CHSS}, under certain circumstances relevant
to the observed quadrupole and octopole, these are closely related to the
maximum-angular-momentum-dispersion-axes of \cite{deOliveira-Costa:2003pu}.
They are also related to the Land and Magueijo triad \cite{LandMagueijoAoE}, at
least for the quadrupole, for which that triad is uniquely defined.  The area
vectors for ILC123 are also to be found in Table \ref{tab:ILC123vectors}.

As expected, the octopole multipole vectors, and consequently the octopole area
vectors, are largely unchanged from ILC1 to ILC123.  Somewhat unexpectedly
(although see \cite{deOliveira-Costa06}) the quadrupole vectors \textit{have}
changed.  This can all be seen quite clearly in Figure
\ref{fig:map:ilc123:2and3}.  To quantify the changes in multipole vectors we
have computed the five dot products between the corresponding year 1 and year
123 quadrupole and octopole multipole vectors
\be
d^{(\ell,i)}\equiv{\hat v}_{\rm ILC1}^{(\ell,i)} \cdot{\hat v}_{\rm ILC123}^{(\ell,i)} .
\ee
Similarly, to quantify the changes in area vectors, we have computed the dot products
between the old and new area vectors
\be
\Delta^{(\ell;i,j)}\equiv \frac{{\vec w}_{\rm ILC1}^{(\ell;i,j)} \cdot{\vec w}_{\rm ILC123}^{(\ell;i,j)}} 
{\vert{\vec w}_{\rm ILC1}^{(\ell;i,j)}\vert\vert {\vec w}_{\rm ILC123}^{(\ell;i,j)}\vert} 
\ee
and also the ratios of lengths between year 1 and year 123 corresponding area vectors
\be
r^{(\ell;i,j)}\equiv \frac{\vert{\vec w}_{\rm ILC1}^{(\ell;i,j)}\vert}{\vert {\vec w}_{\rm ILC123}^{(\ell;i,j)}\vert} .
\ee
These are all found in Table \ref{tab:ILC123dotproducts}, 
together with comparison values for the dot products of vectors in the ILC1 and TOH1 maps.
In the case of the octopole the change from ILC1 to ILC123 is comparable to 
the differences among different full sky maps constructed with the WMAP1
data but the change from ILC1 to ILC123 is considerably larger for the quadrupole.

\begin{figure*}
\includegraphics[width=4in,angle=-90]{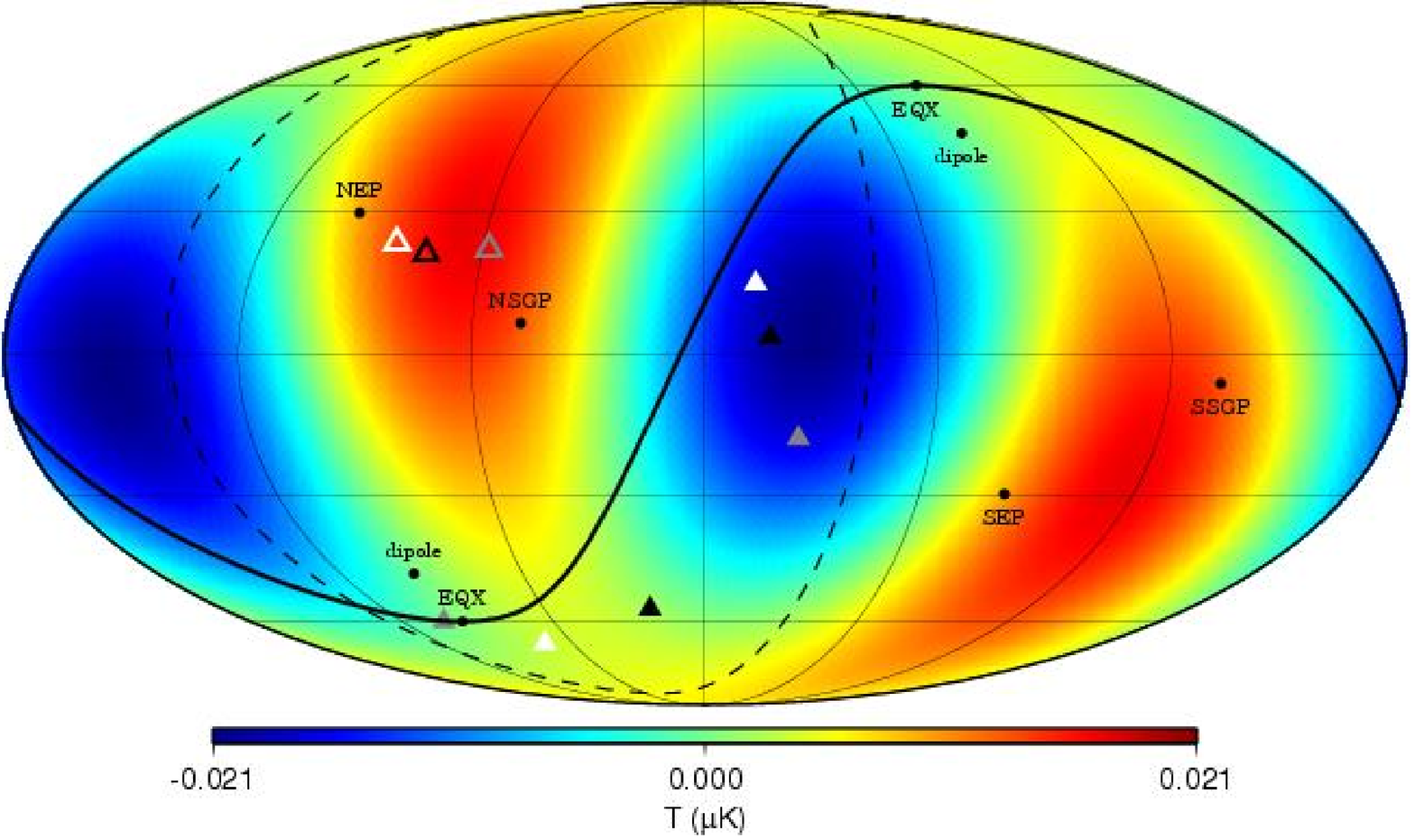}
\caption{The quadrupole vectors (filled triangles) and normals (open triangles)
         of the yr1-yr123 {\it difference} maps for the V band (white), W band 
	 (black) and the ILC (grey). The background shows the ILC123 
	 quadrupole in the same coordinates as in
         Figs.~\ref{fig:map:ilc123:2and3} and \ref{fig:map:ilc123:2+3}.}
        \label{fig:map:diff}
\end{figure*}

The change in the ILC quadrupole is partly due to WMAP's identification of a 
systematic effect in the model of the radiometer gain as a function of time 
\cite{WMAP123_beam} and partly due to the correction of a bias of the ILC map
\cite{WMAP123_angps}.  Let us for the convenience of the reader paraphrase 
the details given in \cite{WMAP123_beam} and \cite{WMAP123_angps} 
to explain both effects. 

The radiometer gain is the voltage difference measured as a result of  
changing sky-temperature differences as the satellite sweeps the sky. 
Besides the sky temperature, the radiometer gain depends on the temperature 
of the optical and electronic components involved, especially the receiver 
box, which houses the radiometers and their electronics. The satellite's 
temperature and with it the temperature of the receiver box changes 
periodically with season and shows a slight long-term increase due to 
the degrading of WMAP's sun shields by the Sun's UV 
radiation \cite{WMAP123_beam}. Consequently, seasonal modulation 
and longer-term temperature drift of the receiver box must be included in a
radiometer gain model, which is needed to calibrate the sky maps.
The gain model is fitted to the hourly measurement of the 
CMB dipole, and consequently the quality of that fit also depends on the 
precision to which the value and orientation of the CMB dipole are known.
While in the calibration of WMAP1 the COBE/DMR dipole has been used,
the WMAP123 analysis relies on the WMAP1 dipole. Having three years of 
data and an improved CMB dipole measurement at hand enabled the WMAP team 
to significantly improve the WMAP123 gain model over the WMAP1 data set. 
The effect of the improved gain model would be expected to be correlated with 
the ecliptic plane, the equinoxes (seasons) and the CMB dipole, due to the reasons given 
above. The seasonal effect must show up most prominently in the quadrupole
(as there are four seasons). 
The WMAP team estimates that the residual error on the quadrupole power is 
$\Delta C_2 = 29 \mu{\rm K}^2$. 

For the synthesis of the ILC (full-sky) map an additional reported
systematic correction (``galaxy bias'') adds to the map a quadrupole and 
octopole aligned with the galaxy to correct for the fact that a minimal 
variance reconstruction of the full sky tends to maximize the cancellation 
between the underlying signal and any foregrounds, which are dominantly 
galactic \cite{WMAP123_angps}.  The WMAP team estimates that the residual 
error of the full-sky ILC after bias correction is $< 5 \mu$K at larger 
than $10$ degrees.  This bias correction apparently affects the overall 
quadrupole power more than the correction to the gain model does, 
but affects the quadrupolar multipole vector structure less.

In particular, we have checked that it is the new gain model that 
slightly changes the orientation of the quadrupole plane 
(defined by $\vec{w}^{(2;1,2)}$) in the individual band maps and in the 
ILC map.  We determined the multipole vectors of the 1yr-123yr difference 
full-sky maps of the V and W bands. As shown in Fig.~\ref{fig:map:diff},
we find that the quadrupole vectors of the difference maps of the bands
are quite close to the ecliptic. The associated normals are 
close both to the ecliptic poles, 
and to one of the maxima of the ILC123 quadrupole 
within the ILC123 quadrupole plane.  (Indeed they lie between them.)
Thus, the 1yr-123yr difference maps for the V and W bands 
show a quadrupole that is orthogonal to the ILC123 quadrupole
and parallel to the ecliptic.
Especially in the V band (supposed to be the least foreground contaminated
band), one of the quadrupole vectors of the difference map not only lies
in the ecliptic plane, but is also very close to the equinoxes. 

\begin{figure*}
	\includegraphics[width=4in,angle=-90]{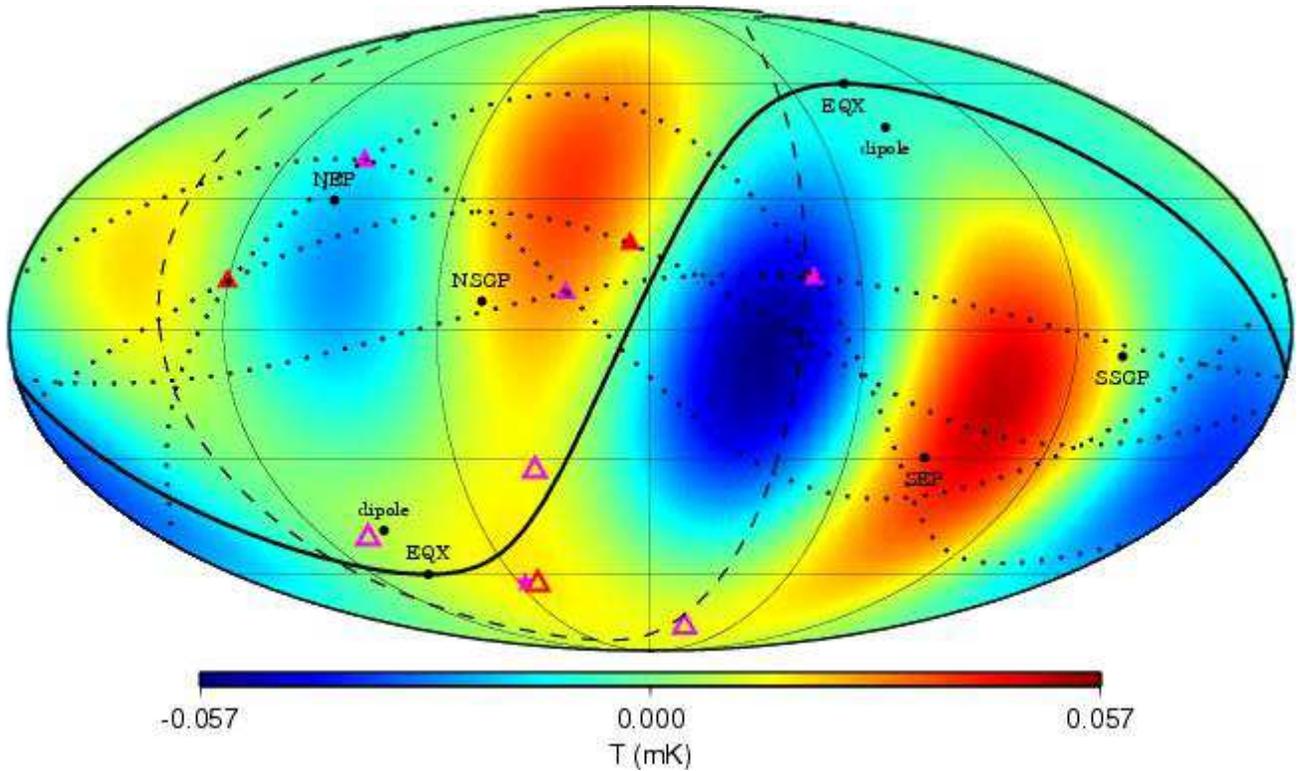}
	\caption{The $\ell=2+3$ multipoles from the ILC123  cleaned map,
	presented in Galactic coordinates.  This is a combination of the 
	two panels of Fig.~\ref{fig:map:ilc123:2and3} with only the
	multipole vectors for the ILC123 map shown for clarity. The solid 
	line is the ecliptic plane and the dashed line is the supergalactic 
	plane.  The directions of the equinoxes (EQX), dipole due to our 
	motion through the Universe, north and south ecliptic poles (NEP and 
	SEP) and north and south supergalactic poles (NSGP and SSGP) are shown.         The $\ell=2$ multipole
	vectors are plotted as the solid red (dark gray in gray scale version)
	triangle and their normal is the open red (dark gray in the gray scale
	version) triangle.  The $\ell=3$ multipole vectors are the solid magenta
	(medium gray in gray scale version) triangles and their three normals 
	are
	the open magenta (medium gray in the gray scale version) triangles.  The
	dotted lines are the great circles connecting the multipole vectors for
	this map (one for the quadrupole vectors and three for the octopole
	vectors).
	The solid magenta (again medium gray in the gray scale version) star 
	is the direction to the maximum angular momentum dispersion for the
	octopole.
        }
	\label{fig:map:ilc123:2+3}
\end{figure*}

The additional effect of the ILC bias correction on the quadrupole 
orientation can be estimated by analyzing the difference map ILC1-ILC123. Now
one of the quadrupole vectors is very close to the equinoxes and the 
quadrupole area is close to the quadrupole areas from the V and W 1yr-123yr 
difference maps. Compared with the difference maps of the individual bands,
the second quadrupole vector is at higher galactic latitude, which is the 
signature of the admixture of a galactic signal from the ILC bias correction
(a galactic signal is expected to have its quadrupole 
vectors at high galactic latitudes \cite{CHSS}). 

WMAP's identification of an ecliptic and equinox 
correlated effect in the radiometer gain is an impressive demonstration 
that the method of multipole vectors is very sensitive to 
systematic effects and/or unexpected physics --- 
and has now proven to be a powerful tool for modern CMB data analysis. 

On the other hand, we think it is quite disturbing that the signature of the 
systematic correction (the difference quadrupole vectors and area) seems to be
anti-correlated (all the vectors are orthogonal to each other) with the 
supposedly cosmic signal. If both the gain model correction
and the bias correction would be all that is needed to get rid of any 
non-cosmic signal, one would expect that the cosmic signal is uncorrelated with the correction. We think that these observations imply the presence of 
rather odd conspiracies between underlying signal and systematics.
\emph{ We suggest that there are additional systematics
or  foregrounds still to be identified.}

\subsection{Multipole  Vectors: Violations of  Statistical Isotropy}
In \cite{SSHC} and \cite{CHSS} we pointed out a number of unusual properties of
the quadrupole and octopole apparent in the WMAP1 full-sky maps (other authors
have analyzed the WMAP sky using the multipole vectors, e.g.\
\cite{Weeks04,Katz2004,SS,Bielewicz05,Abramo06,Helling}.  Here we review these properties for the
ILC123 maps as illustrated in Figures \ref{fig:map:ilc123:2and3} and
\ref{fig:map:ilc123:2+3}.  We refer the reader to Section 4 of \cite{CHSS} for
further details concerning ILC1, TOH1 and LILC1.

\begin{enumerate} 
\item The Queerness of the Quadrupole (cf.\ Fig.\
\ref{fig:map:ilc123:2and3} top panel).  The changes in the quadrupole
multipole vectors are noticeable.
\begin{enumerate} 
\item The normalized quadrupole area vector ${\hat w}^{(2;1,2)}$ 
still lies near the ecliptic plane but has moved slightly away from the plane.
\item ${\hat w}^{(2;1,2)}$ appears somewhat aligned with both the dipole
and the equinoxes but has also moved away from these locations.
\end{enumerate} 
\item The Oddness of the Octopole (cf.\ Fig. \ref{fig:map:ilc123:2and3}
bottom panel). The octopole has not changed beyond expectations.
\begin{enumerate} 
\item Two of the normalized octopole area vectors, ${\hat w}^{(3;1,2)}$
and ${\hat w}^{(3;3,1)}$, lie near the ecliptic plane.
\item The remaining normalized octopole area vector, ${\hat
w}^{(3;2,3)}$, lies very near the supergalactic plane, and about
$10^\circ$ from the Galactic pole.
\item The octopole appears somewhat planar --- the three normalized
octopole area vectors cluster somewhat more than expected.
\end{enumerate} 
\item The Remarkable Relation of the Quadrupole and Octopole
(cf.\ Fig.\ \ref{fig:map:ilc123:2+3}).  For the most part the
relationship of the quadrupole and octopole hasn't changed.
\begin{enumerate} 
\item The normalized quadrupole area vector is aligned with the
normalized octopole area vectors in the sense that it lies ``in the
middle'' of these vectors.  Furthermore, it now lies very close to the
octopole's maximum angular momentum dispersion direction \cite{deOliveira-Costa:2003pu}.
\item The ecliptic carefully traces a zero of the combined map, almost
perfectly over the entire hemisphere containing the Galactic center,
and relatively well over the antipodal hemisphere.
\item Two of the three extrema south of the ecliptic are clearly stronger than
any north of the ecliptic.  The strongest of the three northern extremum is
only comparable in strength to the weakest southern extremum.
\end{enumerate} 
\end{enumerate} 

The above observations are purely qualitative. 
Their statistical significance must be evaluated.  
To this end we developed several statistical tests 
for application to the first-year maps;
we now repeat these tests for ILC123.
Although we have also applied some of these tests to
higher multipoles (with possibly interesting results),
we defer consideration of higher multipoles to a future paper.

\subsubsection{Correlations Among Multipole Vectors}
\label{sec:MVCorrelations}

We first consider the alignment between the quadrupole area vector 
and the three octopole area vectors.  
This alignment can be measured by the magnitudes of the dot products 
between ${\vec w}^{(2;1,2)}$ and each of ${\vec w}^{(3;i,j)}$.
We define these, according to the ordering from the largest to the smallest 
for the first-year full-sky maps (ILC1, TOH1, LILC1):
\bea
A_1 \equiv \vert{\vec w}^{(2;1,2)} \cdot {\vec w}^{(3;1,2)}\vert 
\nonumber \\ 
A_2 \equiv \vert{\vec w}^{(2;1,2)} \cdot {\vec w}^{(3;2,3)}\vert \\
A_3 \equiv \vert{\vec w}^{(2;1,2)} \cdot {\vec w}^{(3;3,1)}\vert
\nonumber 
\eea
A natural choice of statistic which defines an ordering relation on the 
three dot products is \cite{SSHC, Katz2004, CHSS}
\be
S(A_i) = \frac{1}{3} \sum_{i=1}^3 A_i  .
\ee
In Table \ref{tab:AiandScompare} the values of the $A_i$ and $S(A_i)$ are shown
for ILC1, TOH1, LILC1 and ILC123.  
As expected, the differences between ILC1 and ILC123 appear comparable to the 
systematic errors in making full sky maps, as measured by the differences 
between the ILC1, TOH1 and LILC1 values. However, in contrast to the first 
year maps the ILC123 obeys the ordering $A_2 > A_1 > A_3$, which is due to the 
fact that the area vector of the quadrupole has changed significantly. 

\begin{table}[t]
\caption{ Comparison of the values of the $A_i$ 
(the dot products of the quadrupole area vector with 
each of the three octopole area vectors).
$A_i$ and $S(A_i)\equiv(A_1+A_2+A_3)/3$ are tabulated for ILC1, TOH1, 
LILC1 and ILC123, as well as the percentile rank (out of 100) of $S(A_i)$
among a suite of $10^5$ Gaussian random statistically isotropic skies.
For the ILC123 map we show the results with and without the DQ-correction.
}
\label{tab:AiandScompare}
\begin{tabular}{lccccc}
\hline
& ILC1  & TOH1 & LILC1 & ILC123 & ILC123 \\[-0.1cm] 
&      &      &       & uncorr &    \\
\hline
$A_1$ 	                        &0.9251	&0.8509 &0.7798	& 0.716 & 0.7666\\
$A_2$ 	                        &0.7864	&0.7829	&0.7434	& 0.757 & 0.7924\\
$A_3$ 	                        &0.6077	&0.7616	&0.7229	& 0.626 & 0.7221\\
\hline
$S(A_i)$                        &0.7731	&0.7985	&0.7487	& 0.700 & 0.7604\\
${\cal P}\left[S(A_i)\right]$ 	& 0.289	&0.117	& 0.602 & 1.831 & 0.433	\\
\hline
\end{tabular} 
\end{table} 

We have compared the values of the $A_i$ against $10^5$ Monte Carlos of
Gaussian random statistically isotropic skies with pixel noise (as in
\cite{vectors}).  Of course, WMAP1 data is compared to Monte Carlo simulations
with the one-year pixel noise, while for WMAP123 data we use the appropriate
three-year pixel noise.
The percentile rank ${\cal P}$ of the $S(A_i)$ values for each of the four full
sky maps among the $10^5$ associated Monte Carlos is also listed in
Table~\ref{tab:AiandScompare}.  The
clear inference to be drawn is that the quadrupole and octopole show a
statistically significant correlation with each other in the ILC123 map, just
as they do in the WMAP first-year all-sky maps.  This correlation is at the 
$99.6\%$C.L\@. for ILC123,  within the range  seen for the WMAP1 maps ($99.4-99.9\%$C.L\@). 

Also included in Table \ref{tab:AiandScompare} are the values of the $A_i$ and
the associated statistics for an ILC123 from which the Doppler contribution
to the quadrupole has not been removed (ILC123 uncorr).  As for  WMAP123,
we see that the correlation between quadrupole and octopole is significantly
stronger in the properly corrected map than in the uncorrected map.

\subsubsection{Correlations of area vectors with physical directions}

So far we have looked only at the internal correlations between the CMB
quadrupole and octopole.  We have previously \cite{SSHC,CHSS} examined the
correlation of the quadrupole and octopole area vectors with various physical
directions, and  inferred that the  
quadrupole and octopole are also correlated not just with each other
but also with the geometry of the solar system.  
We now reexamine these correlations quantitatively for WMAP123.

Our measure of alignment of the area vectors with a physical direction 
${\hat d}$ is
\bea
S^{(4,4)}({\hat d}) \equiv & 
\displaystyle {1\over 4} \left (\vert {\hat d}\cdot {\vec w}^{(2;1,2)} \vert 
+\vert {\hat d}\cdot {\vec w}^{(3;1,2)} \vert \right. \\ 
&\left.  +\vert {\hat d}\cdot {\vec w}^{(3;2,3)} \vert 
+\vert {\hat d}\cdot {\vec w}^{(3;3,1)} \vert \right ). \nonumber 
\eea
In Table \ref{tab:S_stat_WMAP123}, we display the probabilities
of the $S^{(4,4)}({\hat d})$  statistic for the various WMAP1 and 
WMAP123 full-sky maps
computed for: the ecliptic plane, the  galactic poles (NGP), the supergalactic plane, the CMB dipole
and the  equinoxes.  
These directions and planes are the ones most obviously connected either to possible
sources of systematics (ecliptic plane, CMB dipole and equinoxes) or to possible
unanticipated foregrounds (ecliptic plane, galactic plane/pole, supergalactic plane).
For the galactic poles, the dipole and the equinoxes, these probabilities
are the percentage of statistically isotropic skies 
(among our usual suite of $10^5$  Monte Carlo realizations of statistically isotropic skies)
that exhibit a higher value of $S^{(4,4)}$ for that direction;
for the ecliptic and  supergalactic planes, 
they are the percentage of statistically isotropic skies 
(among the $10^5$  realizations)
that exhibit a lower value of $S^{(4,4)}$ for the axis of each plane.

\begin{table}
\caption{The percentiles of $S^{(4,4)}(\hat{d})$ from $10^5$ MC maps 
test for the alignment of known directions with the quadrupole and 
octopole area vectors. The ILC1, TOH1, LILC1 and ILC123 maps are studied. 
For the ILC123 map we show the results with and without DQ-correction.}
\label{tab:S_stat_WMAP123}
\begin{tabular}[b]{lccccc}
\hline
 & ILC1 & TOH1 & LILC1 & ILC123 & ILC123 \\[-0.1cm] 
 &      &      &       & uncorr &   \\
\hline
ecliptic & 2.01 & 1.43 & 1.48 & 4.88 & 4.11 \\
NGP      & 0.51 & 0.73 & 0.94 & 1.08 & 0.89 \\
SG plane & 8.9  & 14.4 & 13.4 & 10.8 & 15.3 \\
dipole   & 0.110 & 0.045 & 0.214 & 0.489 & 0.269 \\
equinox  & 0.055 & 0.031 & 0.167 & 0.328 & 0.194 \\
\hline
\end{tabular}
\end{table}

From Table \ref{tab:S_stat_WMAP123}, we see that the correlations of
the full sky map to the dipole, equinox and  galactic poles remains
essentially unchanged in the WMAP123 at $99.7\%$C.L\@., $99.8\%$C.L\@. and
$99\%$C.L\@. respectively.  Similarly the correlation to the supergalactic
plane remains insignificant at $85\%$C.L\@.  The correlation with the ecliptic
has declined somewhat from $98\%$C.L\@. to $96\%$C.L\@.

In Section \ref{sec:MVCorrelations} we established the correlation
of the quadrupole and octopole area vectors with each other.  One might
therefore ask whether the observed correlations to physical directions could be
mere accidents of the internal quadrupole-octopole correlations.  In
\cite{CHSS}, we showed that for the first-year maps, even given the ``shape''
of the quadrupole and octopole -- their multipole structure and mutual
orientation -- the additional correlations of their area vectors to the
ecliptic plane were very significant. We also found that the additional
correlations to the dipole and equinoxes was less significant, and the
additional correlations to the galaxy and to the supergalactic plane were not
statistically significant.  We next revisit the issue for the ILC123.

To address this question, we compute the probability, {\it given} the shape of
the quadrupole and octopole and their mutual alignment, that they would align
to the extent they do with each direction or plane.  Put another way, we find
the fraction of directions/planes that are better aligned with the observed
quadrupole and octopole than each of the directions and planes in question.
Thus, we hold the quadrupole and octopole area vectors fixed and compute
$S^{(4,4)}({\hat d})$ both for the physical directions that may be of interest
--- the ecliptic poles, the Galactic poles, the supergalactic poles, the
cosmological dipole and the equinoxes --- and for a large number of random
directions. We denote this test by $S^{(4,4)}({\hat d}|\{\vec{w}\})$, meaning
that the area vectors $\{\vec{w}\}$ are fixed to be the observed ones.

\begin{figure}
\includegraphics[width=2.8in,angle=-90]{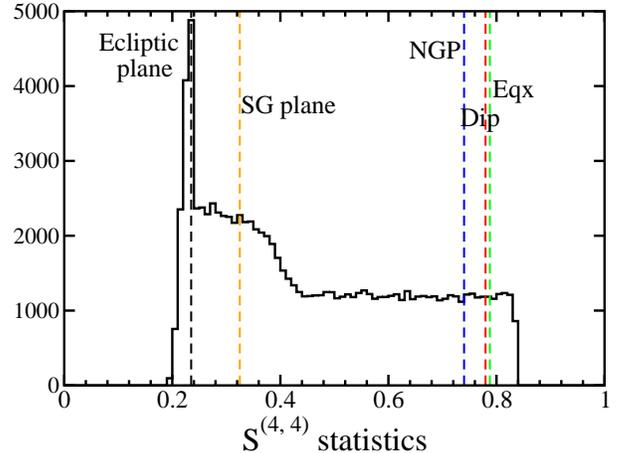}\hspace{-0.5cm}
\caption{Histogram of the $S^{(4, 4)}({\hat d}|\{\vec{w}\})$ statistics
applied to the ILC123 map quadrupole and octopole area vectors and a fixed
direction or plane on the sky, compared to $10^5$ random directions.
Vertical lines show the $S$ statistics of the actual area vectors applied to
the ecliptic plane, NGP, supergalactic plane, dipole and equinox directions
(Table \ref{tab:given23} shows the actual product percentile ranks among the
random rotations for ILC1, ILC123 and TOH1). This Figure and
Table~\ref{tab:given23} show that, {\it given} the relative location of the
quadrupole-octopole area vectors (i.e.\ their mutual alignment), the dipole
and equinox alignments remain unlikely at about $95\%$ C.L., whereas the
ecliptic alignment, significant at 98.3-99.8\% C.L.\ in year 1, is only
significant at the 90\% C.L.\ in year 123.  The galactic plane and
supergalactic plane alignments remain not significant.  }
\label{fig:given23}
\end{figure}

The results are shown graphically in Figure \ref{fig:given23}, which is a
histogram of the values of $S^{(4,4)}({\hat d}|\{\vec{w}\})$ for the randomly
chosen directions, with the values of $S^{(4,4)}({\hat d}|\{\vec{w}\})$ for the
particular physical directions shown. The interesting shape of the histogram,
including the spike, is dictated by the geometry of the particular mutual orientation of the
quadrupole and octopole. Directions lying within the relatively small  triangle bounded
by the three octopole area vectors -- which contains the quadrupole area vector --
contribute to the spike.  The existence of the spike is therefore a consequence
of quadrupole-octopole alignment.  

It is clear from Figure \ref{fig:given23} that the
additional correlations with the galaxy and with the supergalactic plane are
not particularly significant.  To determine the precise significance we compute
for each physical direction, the fraction of random directions that yield a
larger value of $S^{(4,4)}({\hat d}|\{\vec{w}\})$.  These are given in Table
\ref{tab:given23}.

\begin{table}[t]
\caption{ The percentiles of $S^{(4,4)}({\hat d}|\{\vec{w}\})$ for the 
ecliptic plane, NGP, supergalactic plane, dipole and equinox axes
among a comparison group of $10^5$ random directions (or equivalently
random planes) on the sky.  
The ranks are given for the ILC1, TOH1, LILC1 and ILC123 maps.}
\label{tab:given23}
\begin{tabular}{lcccc}
& ILC1 	& TOH1 & LILC1 & ILC123 \\ 
\hline
ecliptic  		&1.7	&1.0	& 0.2	&10.3	\\
NGP		 	&90	&87	& 88	&88	\\
SG plane         	&25	&34	& 33	&32	\\
dipole 			&94.5	&95.6	& 93.8	&93.0	\\
equinox 		&96.4	&96.1	& 94.4	&94.0	\\
\hline
\end{tabular} 
\end{table} 

In Table \ref{tab:given23} we compare the results obtained from ILC123 to those
obtained from ILC1, TOH1 and LILC1.  
As expected, the dipolar and equinoctic values for $S^{(4,4)}({\hat d}|\{\vec{w}\})$ remain essentially
unchanged, in the 93th percentile and 94th percentile  respectively among all directions.  
However, one should probably interpret this as just a 86\% C.L\@.\ and 88\%C.L\@.  case 
respectively for additional correlation, because we would have
found it equally surprising if they had been in the 7th or 6th percentile.  
The 88th percentile rank of the Galactic polar axis similarly offers only a 76\%
C.L.\ case for additional correlation, while the supergalactic plane is clearly not
additionally  correlated by this measure.   

Our first real surprise is that the rank $S^{(4,4)}({\hat d}|\{\vec{w}\})$ for
the (normal to the) ecliptic plane has increased from the $0.2-1.7$ percentile
range of the three WMAP1 maps to $10$ percentile for the ILC123.  This is
because, although the altered quadrupole caused only a small increase in the
actual value of $S^{(4,4)}({\hat d}|\{\vec{w}\})$ for the dipole, the value
moved from below the peak at low $S^{(4,4)}$ seen in Figure \ref{fig:given23},
into the main body of the peak.  
This still represents $90\%$ C.L.\ evidence
of additional correlation of the ecliptic, {\em not} $80\%$ C.L.\ evidence.
That is because, as we shall see in subsection \ref{sec:nodalline}, 
the ILC123 possesses other features
strongly correlated with the ecliptic, not captured by $S^{(4,4)}({\hat
d}|\{\vec{w}\})$ (or even by the area vectors). These features would be impossible
if the ecliptic $S^{(4,4)}({\hat d}|\{\vec{w}\})$ were in the top 10 percentile
of values.

On the basis of the $S^{(4,4)}({\hat d}|\{\vec{w}\})$ statistic alone, while
the additional correlation with the ecliptic is the most significant among the
physical directions and planes which we have considered, it is now of only
$90\%$C.L.\ significance.  However, as we have suggested above, and explore
below, the $S^{(4,4)}({\hat d}|\{\vec{w}\})$ statistic captures only a
part of the strange connection between the quadrupole and octopole and the
ecliptic.

Some measure of caution should also be retained in dismissing
any Galactic, supergalactic, dipole or equinoctic correlations 
just because the additional correlations are not significant.
We do not know whether the correlations with those directions
shown in Table \ref{tab:S_stat_WMAP123} are ``accidental'' consequences of
the internal correlation of the quadrupole and octopole,
or whether the correlation among the quadrupole and octopole area vectors
is an ``accidental'' consequence of their  correlation  to a physical direction.

Finally, we caution the reader in advance that in Section \ref{sec:ACF} we
will call into question some of the  ``improvements'' in the ILC123 over
the ILC1 that have most likely led to the unexpected decline in statistical significance 
of the additional  correlation of the ecliptic with the quadrupole and octopole.

\subsubsection{Correlations with physical directions not captured by area vectors}
\label{sec:nodalline}

The statistics we have studied above do not use all the information in a
multipole.  
The area vectors alone do not contain all this information;
this is obvious for the quadrupole, and true for higher $\ell$ as well.  
Furthermore, we have considered only the dot products
of the area vectors with the physical directions and planes,
this again reduces the portion of information retained.

In Figure \ref{fig:map:ilc123:2+3}, we see very clearly that the ecliptic plane
traces out a locus of zero of the combined quadrupole and octopole over a broad
swath of the sky --- neatly separating a hot spot in the northern sky from a
cold spot in the south.  Indeed, it seems to separate three strong southern
extrema from three weaker northern ones.  This information is not contained in
the $S^{(4,4)}$ statistic since it depends precisely on an extra rotational
degree of freedom about an axis close to the ecliptic plane.
As discussed at some length in \cite{CHSS}, given the observed internal
correlations of the quadrupole and octopole area vectors and given the
correlation of those area vectors with the ecliptic, the probability of the
ecliptic threading along a zero curve in the way it does is difficult to
quantify exactly, but is certainly less than about $5\%$.  
Furthermore, this correlation could not occur if the area vectors were pointing toward the
ecliptic poles instead of perpendicular to them.  (Hence our assertion that the
probability of the observed area vectors' correlation with the ecliptic is 10\%,
not twice that.)  This remains as true for the ILC123 maps as it was for the
WMAP1 full sky maps.

\subsubsection{Summary}

To conclude our re-analysis of the quadrupole and octopole multipole vectors,
there is still strong evidence for the quadrupole-octopole 
alignment and/or the alignment of the quadrupole and octopole with the 
ecliptic/equinox and dipole. {\em The data do not support the statistical 
isotropy of the sky at the largest angular scales.}

\section{Angular Correlation Function}
\label{sec:ACF}

The near vanishing of the two-point angular correlation function at angular
separations greater than about 60 degrees has been the longest established
among the anomalies in the CMB \cite{DMR4}.  Here we compare the results from 
the one-year WMAP data and the three-year data.  
Since WMAP has not provided an analysis of the angular
correlation function for the year 123 data we provide our own.

\subsection{What is the ``angular correlation function''?}

There are several distinct quantities that are commonly referred to as the
``two point angular correlation function'' and the related ``angular power 
spectrum''.
Here we clarify how the various
definitions depend on the assumptions about statistical isotropy and sky
coverage of the map.

We have first 
\be
\label{eqn:ctilcal}
{\tilde {\cal C}}({\hat e}_1,{\hat e}_2) \equiv \langle T({\hat e}_1)T({\hat e}_2)\rangle
\ee
where $\langle\cdot\rangle$ represents an ensemble average over the temperatures
$T({\hat e}_1)$ and $T({\hat e}_2)$ in the particular directions 
${\hat e}_1$ and ${\hat e}_2$.  
(The precise nature of the ensemble when we have only one universe to observe
is philosophically troubling to many, but will be ignored.)
If the sky  were statistically isotropic, then 
${\tilde {\cal C}}({\hat e}_1,{\hat e}_2)$ depends only on 
the angle $\theta$ between ${\hat e}_1$ and ${\hat e}_2$.
This allows us to define our first two-point angular correlation function
\be
\label{eqn:ctilcaltheta}
{\tilde {\cal C}}(\theta) \equiv \langle T({\hat e}_1)T({\hat e}_2)\rangle_\theta,
\ee
where $\langle\cdot\rangle_\theta$ now represents an ensemble average over 
the temperatures $T({\hat e}_1)$ and $T({\hat e}_2)$ in all pairs of 
directions ${\hat e}_1$ and ${\hat e}_2$ separated by the angle $\theta$. 
The average $\langle\cdot\rangle_\theta$ may stand for the full sky or 
just a finite region of the sky. In the case of statistical isotropy  
any region of the sky provides the same answer,
however, in the case of a statistically anisotropic sky, different regions 
of the sky give rise to different values of ${\tilde {\cal C}}(\theta)$.

We can represent ${\tilde {\cal C}}(\theta)$, as a sum involving the Legendre polynomials:
\begin{eqnarray}
\label{eqn:ctilcalthetaLeg}
{\tilde {\cal C}}(\theta) &=& \frac{1}{4\pi} \sum_{\ell=0}^\infty (2\ell+1) 
{\tilde {\cal C}}_\ell P_\ell(\cos\theta), \\
{\tilde {\cal C}}_\ell &\equiv & 2\pi \int_{-1}^1 d (\cos\theta) 
{\tilde {\cal C}}(\theta) P_\ell(\cos\theta).
\end{eqnarray}
These ${\tilde {\cal C}}_\ell$ are also our first incarnation of the 
``angular power spectrum''.

The two point angular correlation function of equation (\ref{eqn:ctilcaltheta})
is defined as an ensemble average.  
By definition,
we can get an unbiased estimate ${\cal C}(\theta)$  
of ${\tilde {\cal C}}(\theta)$
by replacing the ensemble average (for the observed part of the sky)
with an average over the observed sky, even if this is not the full sky:
\be
\label{eqn:ccaltheta}
{\cal C}(\theta) \equiv {\overline{ T({\hat e}_1)T({\hat e}_2)}}_\theta.
\ee
This average is over all pairs of directions
${\hat e}_1$ and ${\hat e}_2$ separated by the angle $\theta$
on the actual sky.
The more, independent pixels separated by an angle $\theta$  
measured on the sky, the more accurate the estimate; but even if the 
sky is cut, the estimate is unbiased.
${ \cal C}(\theta)$ can also be written as a sum over Legendre polynomials:
\be
\label{eqn:ccalthetaLeg}
{ \cal C}(\theta) = \frac{1}{4\pi} \sum_{\ell=0}^\infty (2\ell+1) 
{ \cal C}_\ell P_\ell(\cos\theta).
\ee

An alternative to performing the average over the observed sky
and then a Legendre polynomial expansion to obtain the ${\cal C}_\ell$
is to do a maximum likelihood estimate  (MLE) 
to obtain ${\cal C}^\mathrm{MLE}_\ell$. 
This has several advantages. It allows one to account for noise; 
for the prior probability distribution  of the ${\cal C}_\ell$
(in a statistically isotropic model for the origin of fluctuations, such as inflation);
and for residual galactic uncertainty \cite{slosar_a}
(by marginalizing over foreground templates).
Because the MLE method gives a distribution of likelihoods as a function of the 
${\cal C}^\mathrm{MLE}_\ell$, one has a measure not just of the most likely value
of each ${\cal C}_\ell$ but also of the uncertainty in that value.

Some caution is necessary in using  the MLE.  
MLE assumes that each $T({\hat e}_i)$  is Gaussian distributed.
It also assumes statistically isotropic noise.
For $\ell\leq10$, the WMAP team quotes \cite{WMAP123_angps} 
the peak in the likelihood distribution of 
${\cal C}^\mathrm{MLE}_\ell$
as the value of  the ``angular power spectrum'' for that $\ell$.

There is another closely related quantity, which is often called the two-point
angular correlation function, but which is not identical to the ensemble
average~(\ref{eqn:ctilcaltheta}) except in certain circumstances.
Consider the usual spherical harmonic expansion of the CMB sky 
\be
T \left(\theta,\phi\right) \equiv 
\sum_{\ell=0}^{\infty} T_\ell \equiv 
\sum_{\ell=0}^{\infty}\sum_{m=-\ell}^{\ell} a_{\ell m} 
Y_{\ell m}\left(\theta,\phi\right).
\ee
The ensemble average of the product of $a_{\ell m}$,
\be
{\tilde C}_{\ell m \ell' m'} \equiv \langle a_{\ell m}^* a_{\ell' m'} \rangle ,
\ee
encodes the same information as ${\tilde C}({\hat e}_1,{\hat e}_2)$.
The sky is statistically isotropic if and only if
\be
\label{eqn:SI}
{\tilde C}_{\ell m \ell' m'} 
= {\tilde C}_\ell \delta_{\ell \ell'}\delta_{m m'}  .
\ee
In this case one can show that the angular power spectrum ${\tilde C}_{\ell}$ 
are identical to the coefficients in equation (\ref{eqn:ctilcalthetaLeg}),
i.e. 
\be 
\label{eqn:SIimplies}
\mbox{statistical isotropy} \Rightarrow
\left({\tilde C}_\ell = {\tilde {\cal C}}_\ell \right) .
\ee
Similarly, one can define a function based on the ${\tilde C}_\ell$,
\be
{ \tilde C}(\theta) \equiv \frac{1}{4\pi} \sum_{\ell=0}^\infty (2\ell+1) 
{ \tilde C}_\ell P_\ell(\cos\theta) . 
\ee

Unfortunately, we cannot compute the ${\tilde C}_\ell$ from what we measure
since the ${\tilde C}_\ell$ are properties of the underlying ensemble 
of which the sky is but one realization. 
What we can compute are estimators of ${\tilde C}_\ell$.
For example, given the full sky, we could measure the $a_{\ell m}$ from the 
full sky and then define 
\be
\label{eqn:Cl}
C_\ell \equiv 
\frac{1}{2\ell+1}\sum_{m=-\ell}^{\ell} \left\vert a_{\ell m} \right\vert^2 .
\ee 
Obviously this definition is motivated by statistical isotropy, but is not 
restricted to that case. In the case of statistical isotropy it is an unbiased 
estimator of $\tilde{C}_\ell$. Additionally, if Gaussianity holds, this is the
best estimator with cosmic variance 
${\rm Var}(C_\ell) = 2 {\tilde C}_\ell^2/(2\ell +1)$. 
If Gaussianity is violated the cosmic variance may be larger. 

The estimator (\ref{eqn:Cl}) can be easily adapted to cut skies.
Other similar estimators exist (and have improved performance, for example
in a noisy environment \cite{WMAP123_angps}).
If $\ell$ is large, $C_\ell$ will be a good estimate of the angular power
spectrum ${\tilde C}_\ell$.
These $C_\ell$, however precisely obtained, are often called the pseudo-$C_\ell$.

Finally, one can define the function based on the $C_\ell$,
\be
{  C}(\theta) \equiv \frac{1}{4\pi} \sum_{\ell=0}^\infty (2\ell+1) 
{  C}_\ell P_\ell(\cos\theta)  .
\ee
In the case of statistical isotropy and Gaussianity the cosmic variance of the 
correlation function estimated from full skies becomes 
\begin{equation}
{\rm Var}(C(\theta)) = \frac{1}{8\pi^2}\sum_{\ell = 0}^\infty (2\ell +1) 
{\tilde C}_\ell^2 P_\ell^2(\cos\theta).
\end{equation}

If the $C_{\ell}$ and ${\cal C}(\theta)$ are inferred from a full sky
measurement, then one can show that (up to pixelization errors)
\be
\label{eqn:fullskyimplies1}
\mbox{full sky} \Rightarrow 
\left({C}_\ell = {\cal C}_\ell \right) ,
\ee
and equivalently
\be
\label{eqn:fullskyimplies2}
\mbox{full sky} \Rightarrow 
\left({C}(\theta) = {\cal C}(\theta) \right) .
\ee
The relationships among these quantities are summarized in Table \ref{tab:Cs}.

While the temptation not to differentiate between these quantities is
understandable, we would argue for a preference to reserve for ${\tilde{ \cal
C}}(\theta)$ (and ${ \cal C}(\theta)$) the name angular correlation function
(and its estimator); and to call ${\tilde C}_\ell$ (and $C_\ell$) the angular
power spectrum (and its estimator).

\begin{table}[t]
\caption{ The four classes of pairs of quantities often called the ``two-point
angular correlation function'' and the ``angular power spectrum''.  Angle
brackets represent ensemble averages, an overbar represents an average over the
sky or some portion thereof.  LPT stands for Legendre polynomial transform, LPS
for Legendre polynomial series expansion.  $T_i$ is the temperature in the
$i$th sky pixel, $a_{\ell m}$ is the coefficient of $Y_{\ell m}$ in a spherical
harmonic expansion of the temperature field on the sky.  In the case of
$C_\ell$, the definition actually represents the canonical, but not necessarily
best, estimator of ${\tilde C}_\ell$, calculated from the spherical harmonic
coefficients of the temperature field.  }
\label{tab:Cs}
\begin{tabular}{cccc}
\hline
``correlation&function'' & ``angular & power-spectrum'' \\ 
\hline
symbol     & definition     & symbol      & definition  \\
\hline
${\tilde {\cal C}}(\theta)$& $\langle T_i T_j \rangle$ &
${\tilde {\cal C}}_\ell	  $& LPS of ${\tilde {\cal C}}(\theta)  $ \\
$	 {\cal C} (\theta)$& ${\overline { T_i T_j }}     $ &
$	 {\cal C} _\ell	  $& LPS of        ${\cal C} (\theta)   $  \\
${\tilde       C} (\theta)$& LPT of ${\tilde C}_\ell		$ &
${\tilde       C} _\ell	  $& $\langle a_{\ell m}^* a_{\ell m} \rangle $ \\
$              C  (\theta)$& LPT of         $C_\ell		$ &
$              C  _\ell	  $& $\frac{1}{(2\ell+1)} 
			   \sum_m\vert a_{\ell m}\vert^2 $ \\
\hline
\end{tabular} 
\end{table} 

In the previous Section we presented evidence that the low-$\ell$ sky 
is {\em not} statistically isotropic. Moreover the $a_{\ell m}$
(or the $C_\ell$)
are generically {\em not} computed over a full sky. Therefore the
distinctions between 
${\tilde {\cal C}}$ and  ${\tilde C}$, 
and  between ${\cal C}$ and $C$
must be kept in mind when interpreting CMB results.  In particular,
the measurable quantities --- ${\cal C}(\theta)$ and $C_\ell$ --- are
{\em not} interchangeable, but rather encode, at least in principle
somewhat {\em different} information about the  map.  This means
that one must check that any analysis procedure has reasonable
consequences not just for the $C_\ell$, but also for ${\cal C}(\theta)$.

Indeed, in the absence of statistical isotropy, 
many of the quantities  to which we are accustomed  are of uncertain meaning
and dubious value.  For example, neither ${\tilde {\cal C}}(\theta)$ nor  ${\tilde C}_\ell$
characterizes the statistical properties of the ensemble 
of which the CMB is a realization.  
They are averages of quantities that may (partly) characterize the ensemble ---
${\tilde C}({\hat e}_1,{\hat e}_2)$ or ${\tilde C}_{\ell m \ell' m'}$  ---
but, in the absence of SI,
these more complicated quantities cannot be meaningfully estimated with only one sky.
Estimators like 
${\cal C}(\theta)$ and $C_\ell$ remain of interest primarily to the extent that either
the sky is almost statistically isotropic (so that 
${\tilde C}({\hat e}_1,{\hat e}_2)$ and ${\tilde C}_{\ell m \ell' m'}$  
can be expanded around 
${\tilde {\cal C}}(\theta)$ and ${\tilde C}_\ell\delta_{\ell\ell'}\delta_{m m'}$)
or the estimators (${\cal C}(\theta)$ or $C_\ell$) return unexpected values.

The question of approximate statistical isotropy may be worth exploring.  
What is certainly true, as we show below,  is that $\vert {\cal C}(\theta)\vert$  
remains unexpectedly small at large angles in WMAP123, just as it was for COBE-DMR and WMAP1.
One must therefore be extremely cautious about any reconstruction 
method that regards the ${\tilde C}_\ell$ as fundamental, independent
variables, while ignoring the properties of ${\cal C}(\theta)$.

To summarize, while CMB cosmologists have a long-standing interest in and
preference for the angular power spectrum ${\tilde C}_\ell$ as the relevant
variables for statistical analysis, given the absence of statistical isotropy
for low $\ell$ these variables are at best flawed.  Therefore one must be
careful to distinguish between ${\tilde C}_{\ell}$ and ${\tilde {\cal C}}_\ell$
and their estimators; one must be cautious that ${\tilde C}_{\ell}$ and
${\tilde {\cal C}}_\ell$ may not characterize 
the observed microwave maps;
and one must pay careful attention to the properties of ${\cal C}(\theta)$.

\subsection{The measured ${\cal C}(\theta)$}

\begin{figure*}
\includegraphics[width=2.8in,angle=-90]{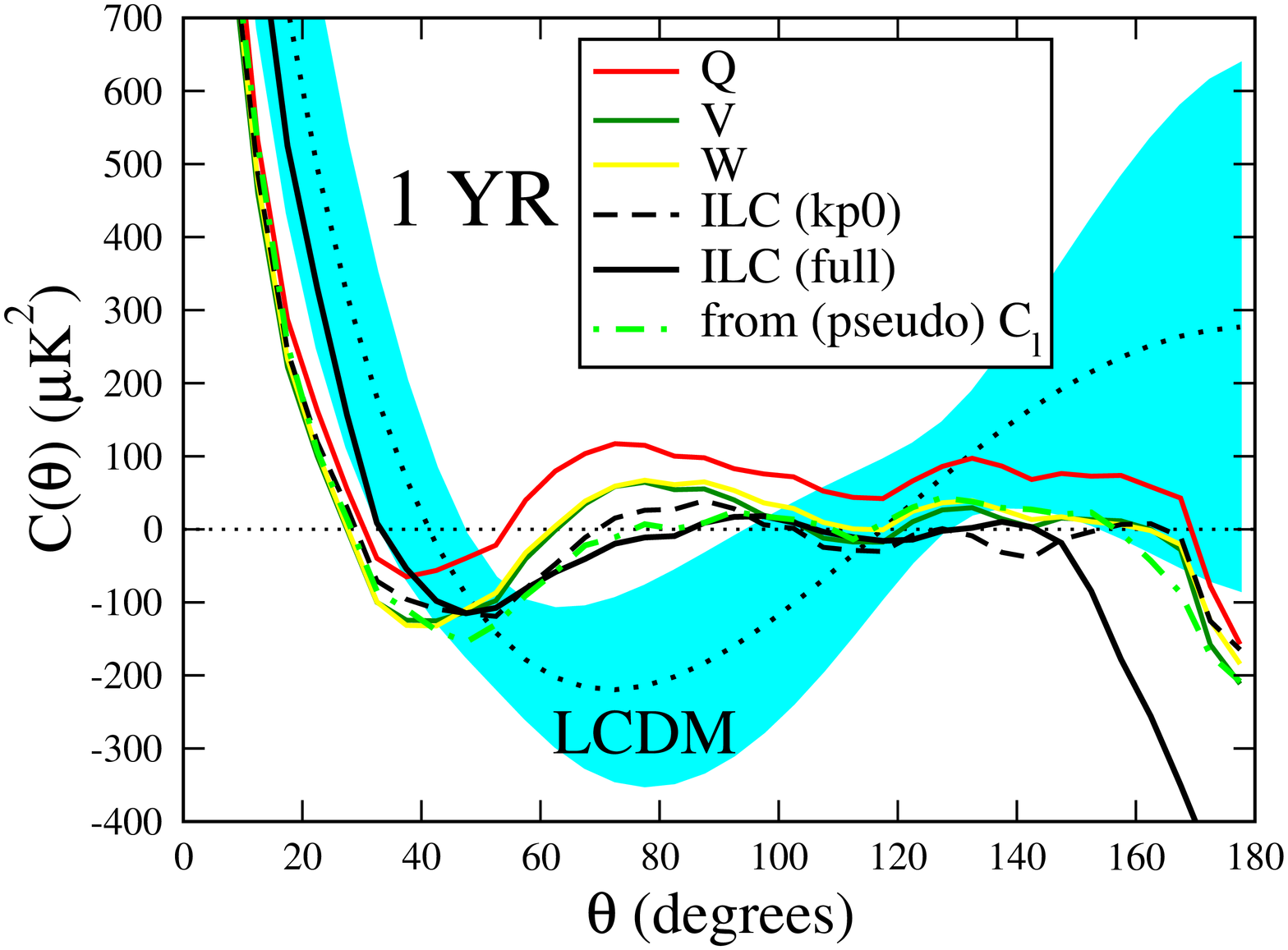}\hspace{-0.75cm}
\includegraphics[width=2.8in,angle=-90]{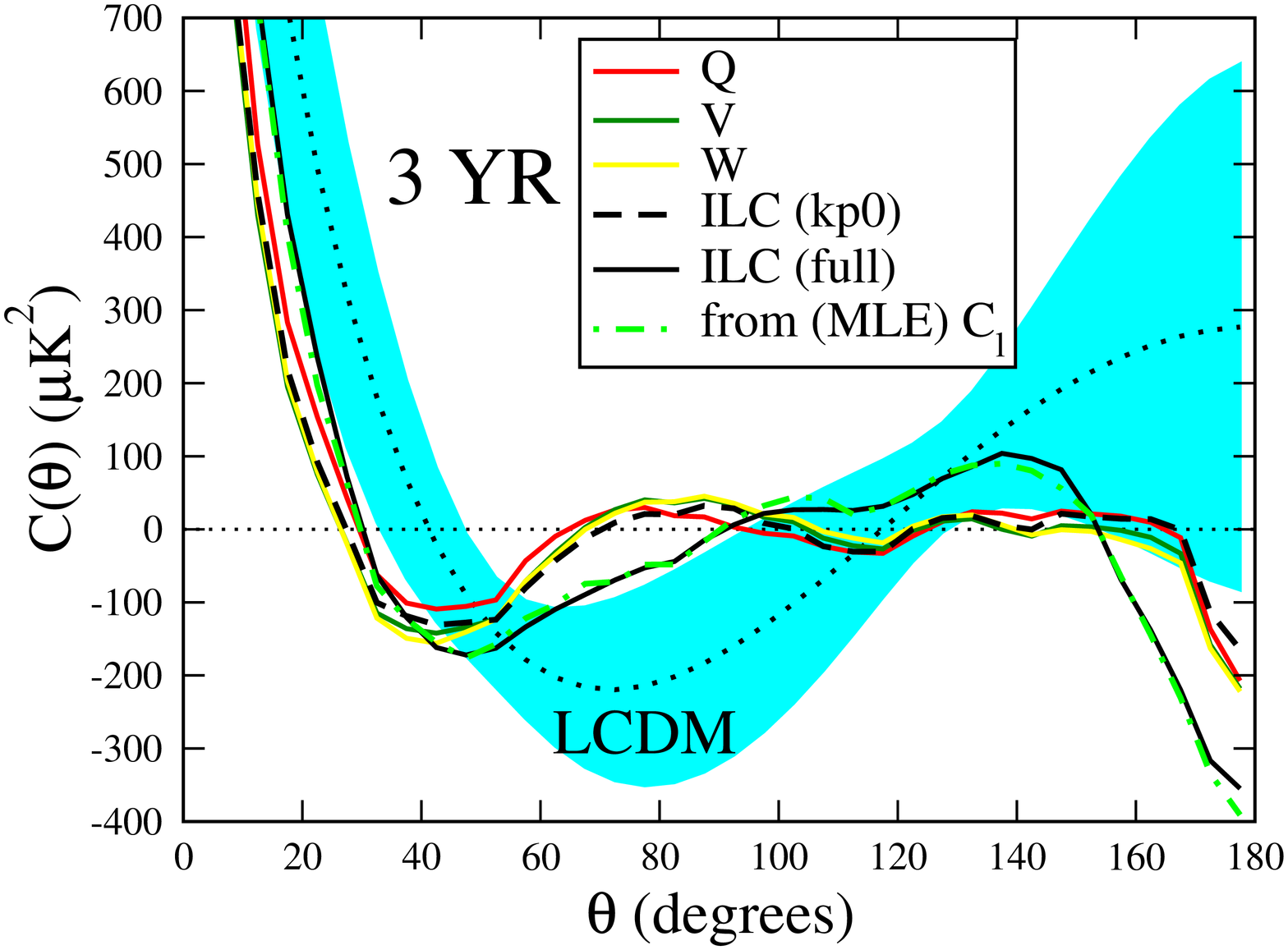}
\caption{Two point angular correlation function, ${\cal C}(\theta) \equiv
{\overline{ T({\hat e}_1)T({\hat e}_2)}}_\theta$ computed in pixel
space, for three different bands masked with the kp0 mask. Also shown is the
correlation function for the ILC map with and without the mask, and the value
expected for a statistically isotropic sky with best-fit $\Lambda$CDM cosmology
together with 68\% error bars.  Left panel: year 1 results.  Right panel: year
123 results. Even by eye, it is apparent that masked year 123 maps have 
$C(\theta)$   that is consistent with zero at $\theta\gtrsim 60$ deg, even more so
than in year 1 maps. We also show the $C(\theta)$ computed from the ``official'' published
$C_\ell$, which (at $\ell<10$) are the pseudo-$C_\ell$ in year 1, the and MLE $C_\ell$
in year 123. Clearly, the MLE-based $C_\ell$, as well as ${\cal C}(\theta)$
computed from the full-sky ILC maps, are in significant disagreement with the angular
correlation function computed from cut-sky maps.}
\label{fig:ctheta}
\end{figure*}

We can readily explore ${\cal C}(\theta)$, the angular correlation function,
since there is indeed a very large number of independently measured pixels on the
WMAP sky.  ${\cal C}(\theta)$ was first measured (at large angles) using the
Cosmic Background Explorer's Differential Microwave Radiometer (COBE-DMR)
\cite{DMR4}, and found to be anomalously small in magnitude at large angles in
all COBE bands.

The WMAP team \cite{WMAP1_results,WMAP1_low} using their first year of data
presented the ``angular correlation function'', but this was obtained from the
$C_\ell$ (and so, was actually $C(\theta)$, not ${\cal C}(\theta)$).  WMAP did
not present any angular correlation function in their third-year papers,
although, as mentioned above, their ``angular power spectrum'' at low $\ell$
were actually maximum likelihood estimates of ${\cal C}_\ell$, i.e. the
coefficients of a Legendre polynomial series expansion of a MLE of ${\cal
C}(\theta)$.

In Figure \ref{fig:ctheta} we show ${\cal C}(\theta) \equiv {\overline{ T({\hat
e}_1)T({\hat e}_2)}}_\theta$ (computed in pixel space) for the Q, V and W band maps
masked with the kp0 mask. (See \cite{WMAP1_results} for a discussion of the various
galactic masks.) The K and Ka bands have a large and obvious galactic
contamination even after applying the kp0 mask, and we do not use them.  Also
shown are the correlation function for the ILC map with and without the mask,
and the Legendre polynomial series of the quoted maximum likelihood estimates of ${\cal C}_\ell$,
i.e. ${\cal C}^\mathrm{MLE}(\theta)$.
Finally, we show ${\tilde {\cal C}}(\theta)={\tilde C}(\theta)$ for the best
fit $\Lambda$CDM model of the three-year WMAP, and a blue band around that
showing the range of $C(\theta)$ that one would extract from a full sky, given
cosmic variance. The left panel shows the year 1 results, while the right panel
shows year 123 results.

First of all, we see that the different bands and the cut-sky ILC map are all
in excellent agreement with each other for year 1 and year 123 separately,
perhaps with an exception of the year 1 Q band map. In each case the magnitude
of ${\cal C}(\theta)$ is very small above $60$ degrees and below $170$ degrees;
above $170$ degrees there is a peculiar slight anti-correlation while the model
predicts a large positive correlation. Moreover, even by eye it is apparent
that year 123 maps have ${\cal C}(\theta)$ that is nearly vanishing at
$60\lesssim \theta\lesssim 170$ deg even more precisely than in year 1
maps. Finally, the full-sky ${\cal C}(\theta)$ from the ILC maps is in
agreement with i.e. ${\cal C}^\mathrm{MLE}(\theta)$, but in significant
disagreement with the cut-sky value.

We now compare the smallness of ${\cal C}(\theta)$ to that expected in simulated maps.
We use the statistic proposed by WMAP \cite{WMAP1_low} in their first-year papers
that quantifies the lack of power at $\theta>60$ degrees
\be
\label{eqn:Shalf}
S_{1/2} \equiv \int_{-1}^{1/2} \left[ C(\theta)\right]^2 d (\cos\theta).
\ee

The WMAP team computed
$S_{1/2}$ for the original ILC1 and found that only $0.15\%$ of the elements in
their Markov chain of $\Lambda$CDM model CMB skies had lower values of
$S_{1/2}$ than the true sky.  In addition we define the statistic
\be
S_{\rm full} \equiv \int_{-1}^{1} \left[ C(\theta)\right]^2 d (\cos\theta)  ,
\ee
which quantifies the amount of power squared on {\it all} angular scales.

In Table \ref{tab:Shalf} we give the value of $S_{1/2}$ for  the one-year and three-year
band maps, for ILC1 and ILC123, and for the best fit $\Lambda$CDM model.
The values of $S_{1/2}$ and $S_{\rm full}$ are very low relative to what the
$\Lambda$CDM isotropic cosmology predicts and are noticeably lower in the year 123
maps than in the year 1 maps.  

\begin{table*}[t]
\caption{ Values of $S_{1/2}$ and $S_{\rm full}$ statistics, and their ranks
(in percent) relative to Monte Carlo realizations of isotropic skies consistent
with best-fit $\Lambda$CDM models from WMAP 3 year constraints. All of the maps
were subjected to the kp0 mask and compared to 20,000 Monte Carlo realizations
with the kp0 mask also applied. The third and second-to-last row are without
the mask and compared to 1000 full-sky simulations, while the last row
shows the $\Lambda$CDM expectation. The boldfaced entries show
the extremely small values of $S_{1/2}$ statistic for the 3-year cut-sky maps,
while the single italic entry shows that the full-sky ILC123 map does {\it not}
have an unusually small ${\cal C}(\theta)$.  We also show the quadrupole,
octopole and hexadecapole power ${\cal C}_\ell$ estimated from the
corresponding ${\cal C}(\theta)$; note again the difference between the cut-sky
and full-sky values.  }
\label{tab:Shalf}
\begin{tabular}{cccccccc}
\hline
Map & $S_{1/2}$ & $P(S_{1/2})$ & $S_{\rm full}$ & $P(S_{\rm full})$ 
&  $6\,{\cal C}_2/(2\pi)$ &$12\,{\cal C}_3/(2\pi)$ &$20\,{\cal C}_4/(2\pi)$  \\
& $(\mu K)^2$ &$(\%)$ &$(\mu K)^2$ &$(\%)$
& $(\mu K)^2$ &$(\mu K)^2$ &$(\mu K)^2$ \\
\hline
Q123          &  956 & {\bf 0.04} & 33399 & 1.91  & 178  & 430  &799\\
V123          & 1306 & {\bf 0.12} & 27816 & 0.96  & 47   & 403  &818\\
W123          & 1374 & {\bf 0.14} & 29706 & 1.27  & 35   & 449  &828\\
Q1            & 10471& 10.2       & 41499 & 4.3   & 79   & 357  &745\\
V1            & 2117 & 0.38       & 29170 & 1.16  & 53   & 369  &796\\
W1            & 2545 & 0.64       & 30872 & 1.38  & 40   & 405  &815\\ 
\hline
ILC123(cut)   &  928 & {\bf 0.03} & 28141 & 1.04  & 113  & 414  &831\\
ILC1(cut)     & 1172 & 0.09       & 29873 & 1.25  & 115  & 443  &901\\ 
\hline
ILC123(full)   & 8328 & {\it 7.0} & 60806 & 9.1   & 247  &1046  &751\\
ILC1(full)     & 9119 &  8.3      & 70299 & 14.4  & 195  &1050  &828\\
\hline
$\Lambda$CDM  & 39700 & ---       & 133000&  ---  &1091  &1020  &958\\
\hline
\end{tabular} 
\end{table*} 

To quantify the probabilities of our statistics, we perform comparison with
20,000 Gaussian random, isotropic realizations of the skies consistent with
models favored by the WMAP data. The angular power spectrum from {\it each} simulated
sky is generated as follows:

\begin{enumerate}
\item pick a model (in order of the highest weight) from WMAP's publicly 
available Markov chains;

\item generate the angular power spectrum in multipole space corresponding
to that model by running the CAMB code \cite{CAMB};

\item generate a map in Healpix \cite{healpix} consistent with the model;

\item apply the kp0 mask; 

\item degrade to NSIDE=32 resolution, then reapply the kp0 mask
also degraded to NSIDE=32; and

\item compute ${\cal C}(\theta)$ directly from the map, using the unmasked pixels.
\end{enumerate}

In addition, and using the same procedure but without masking, we compute
${\cal C}(\theta)$ from 1000 simulated full-sky maps for comparison to full-sky
ILC maps. (Since $P(S_{1/2})$ and $P(S_{\rm full})$ for the full-sky maps are
not as low as for the cut-sky maps, we did not need to go through the
resource-consuming task of generating 20 times more simulations in order to get
good statistics.) We have checked that applying the inverse-noise weighting to
the pixels when degrading the map resolution (O. Dor\'{e}, private
communication) leads to negligible differences in the recovered ${\cal
C}(\theta)$ ($\lesssim 1\mu K^2$ on large scales).  Finally we have checked
that, for the cut-sky case, applying the kp2 mask instead of kp0 gives similar
results.

Results from the comparisons are given in the third and fifth column of
Table~\ref{tab:Shalf}.  They indicate that only 0.04\%-0.14\% of isotropic
$\Lambda$CDM skies (for the Q, V and W band maps) give $S_{1/2}$ that is smaller than
that from WMAP123. Similarly, the masked ILC123 map has $S_{1/2}$ low at the
0.03\% level. Interestingly, even the $S_{\rm full}$ statistic that includes
power on all scales, applied to cut-sky maps, is low at about a 1\% level.
We therefore conclude that {\it the absence of large angle correlations at
scales greater than 60 degrees is, at $>99.85$\%C.L., even more significant in
year 123 than in year 1}.

We also show the quadrupole, octopole and hexadecapole
power inferred from ${\cal C}(\theta)$ of various maps, as well as the
$\Lambda$CDM expected values.  It is clear that the quadrupole and octopole
computed from cut-sky ${\cal C}(\theta)$ are a factor of two to five smaller
than those from the full-sky maps (and of course the latter are another factor
of two to four smaller than the most likely $\Lambda$CDM values).  This result
has been hinted at by the WMAP team \cite{WMAP123_angps} who pointed out that
the MLE estimates of the quadrupole and octopole (as well as $\ell=7$) 
they adopted in year 123 are considerably larger than the pseudo-$C_\ell$
values they had adopted in year 1. Here we find that the pseudo-$C_\ell$
estimates are in a much better agreement with the cut-sky ${\cal
C}(\theta)$ than the MLE values; see Figure \ref{fig:ctheta}.

The current situation seems to us very disturbing.  Standard procedures for
extracting the ``angular power spectrum'' of the CMB uses pseudo-$C_\ell$.  The
recent WMAP analysis included a maximum-likelihood analysis at low $\ell$
(described above) to correct for the effects of a cut sky and to mitigate any
remaining galaxy contamination.  But both the connection between
pseudo-$C_\ell$ and any properties of the underlying ensemble (whatever that
may mean), and the validity of the maximum-likelihood analysis rely on the
assumption that the underlying sky (i.e.\ the ensemble) is statistically
isotropic.  However, we (and others) have already provided strong evidence that
the underlying sky, or at least the all-sky maps, are {\em not} statistically
isotropic.  In the absence of statistical isotropy none of the computable
quantities have any clear connection to the statistical properties of the
underlying ensemble.  Indeed, in the absence of statistical isotropy each
$a_{\ell m}$ could have its own distinct distribution.  Since we have precisely
one sample from each such distribution -- the value of $a_{\ell m}$ on the
actual sky (and even that only for a measured full sky) -- it is impossible to
in any way characterize these distributions from the observation of just one
universe.

We are faced with two choices: either the universe is not statistically
isotropic on large scales (perhaps because it has non-trivial topology), or the
observed violation of statistical isotropy is not cosmological, but rather due 
either to systematic errors, or to unanticipated foregrounds.  In this latter
case, it is imperative to get to the root of the problems at large angles.

The disagreement between the full-sky and cut-sky angular correlation
functions, evident in Figure \ref{fig:ctheta} and Table~\ref{tab:Shalf}, is
both marked and troubling.  The third and second-to-last columns of Table
\ref{tab:Shalf} show that both the quadrupole and the octopole power estimated
from the cut-sky maps are much smaller than those estimated from reconstructed
full-sky ILC maps -- for each multipole, the former is about one half as large
as the latter.  However, it is the the full-sky values which correspond to the
estimated values used in cosmological parameter analyzes.

It is of utmost concern to us that one begins with individual band maps that
all show very little angular correlation at large angles (i.e.  $\vert {\cal
C}(\theta)\vert$ is small), and, as a result of the analysis procedure, one is
led to conclude that the underlying cosmological correlations are much larger.
What is even more disturbing is that the full-sky map making algorithm is
inserting significant extra large angle power into precisely those portions of
the sky where we have the least reliable information.  We are similarly
concerned that the ``galactic bias correction'' of the ILC123, which inserts
extra low-$\ell$ power into the region of the galactic cut, does so on the
basis of the expectations of a model whose assumptions (statistical isotropy
and the particular $\Lambda$CDM power spectrum) are not borne out even after
all the corrections are made.  But even the first-year ILC, which did not have
this bias correction, suffers from a surfeit of power inside the cut.

 Finally, although the maximum likelihood estimates
of the low-$\ell$ $C_\ell$ (actually ${\cal C}_\ell$) quoted by WMAP in their
three-year release appear to be in better agreement with the $\Lambda$CDM model
than are the values one derives from cut-sky individual bands, 
or the cut sky ILC123,  or the values quoted in the one-year WMAP release,
nevertheless, the angular correlation function that the new quoted $C_\ell$ imply 
(as shown in Figure \ref{fig:ctheta}) appears
to be in even worse agreement with the $\Lambda$CDM prediction than all the others.

\section{Conclusions}

We have shown that the ILC123 map, a full sky map derived from the first three
years of WMAP data like its predecessors the ILC1, TOH1 and LILC1 maps, derived
from the first year of WMAP data, shows statistically significant deviations
from the expected Gaussian-random, statistically isotropic sky with a  
generic inflationary spectrum of perturbations.
In particular: 
there is a dramatic lack of angular correlations at angles greater than sixty degrees; 
the octopole is quite planar
with the three octopole planes aligning with the quadrupole plane; 
these planes are perpendicular to the ecliptic plane 
(albeit at reduced significance than in the first-year full-sky maps); 
the ecliptic plane neatly separates two extrema of the
combined $\ell=2$ and $\ell=3$ map, with the strongest extrema to the south of
the ecliptic and the weaker extrema to the north.  The probability that each of
these would happen by chance are $0.03$\% (quoting the cut-sky ILC123 $S_{1/2}$
probability), $0.4$\%, $10$\%, and $<5\%$.  As they are all independent and all
involve primarily the quadrupole and octopole, they represent a $\sim 10^{-8}$
probability chance ``fluke'' in the two largest scale modes.  To quote
\cite{SSHC}: {\it We find it hard to believe that these correlations are just
statistical fluctuations around standard inflationary cosmology's prediction of
statistically isotropic Gaussian random $a_{\ell m}$ [with a nearly scale-free
primordial spectrum].}

What are the consequences and possible explanations of these correlations?
There are several options --- they are statistical flukes, they are
cosmological in origin, they are due to improper subtraction of known
foregrounds, they are due to a previously unexpected foreground, or they are
due to WMAP systematics.  As remarked above it is difficult for us to accept
the occurrence of a $10^{-8}$ unlikely event as a scientific explanation.

A cosmological mechanism could possibly explain the weakness of large angle
correlations, and the alignment of the quadrupole and octopole to each other.
A cosmological explanation must ignore the observed correlations to the solar
system, as there is no chance that the universe knows about the orientation of
the solar system nor vice-versa.  These latter correlations are unlikely at the
level of less than $1$ in $200$ (plus an additional independent $\approx 1/10$
unlikely correlation with the dipole which we have ignored).  This possibility
seems to us contrived and suggests to us that explanations which do not account
for the connection to solar system geometry should be viewed with considerable
skepticism.

In \cite{CHSS}, we showed that the known Galactic foregrounds possess a
multipole vector structure wholly dissimilar to those of the observed
quadrupole and octopole. This argues strongly against any explanation of the
observed quadrupole and octopole in terms of these known Galactic foregrounds.

A number of authors have attempted to explain the observed quadrupole-octopole
correlations in terms of a new foreground
\cite{Moffat05,Frisch05,Gordon05,Vale05,Inoue06,Ghosh}.  (Some of these also
attempted to explain the absence of large angle correlations, for which there
are also other proffered explanations
\cite{Cline,Linde03,Luminet03,Jaffe03,Weeks03,Abramo03,Gordon04}.)  Only one of the
proposals (\cite{Frisch05}) can possibly explain the ecliptic correlations, as
all the others are extra-galactic.  Some do claim to explain the
less-significant dipole correlations.  Difficulties with individual mechanisms
have been discussed by several authors \cite{Rakic06, Hansen05, Ferrer,
Gordon05, Cooray05} (sometimes before the corresponding proposal).
Unfortunately, in each and every case, among possible other deficiencies, the
pattern of fluctuations proposed is inconsistent with the one observed on the
sky.  As remarked above, the quadrupole of the sky is nearly pure $Y_{22}$ in
the frame where the $z$-axis is parallel to $\hat w^{(2,1,2)}$ (or any nearly
equivalent direction), while the octopole is dominantly $Y_{33}$ in the same
frame.  Mechanisms which produce an alteration of the microwave signal from a
relatively small patch of sky --- and all of the above proposals fall into this
class --- are most likely to produce aligned $Y_{20}$ and $Y_{30}$.  (This is
because if there is only one preferred direction, then the multipole vectors of
the affected multipoles will all be parallel to each other, leading to a
$Y_{\ell 0}$.)  The authors of \cite{Inoue06} manage to ameliorate the
situation slightly by constructing a distorted patch, leading to an
under-powered $Y_{33}$, but still a pure $Y_{20}$. The second shortcoming of
all explanations where contaminating effect is effectively {\it added} on top
of intrinsic CMB temperature is that chance cancellation is typically required
to produce the low power at large scales, or else the intrinsic CMB happens to
have {\it even less} power than what we observe.  Likelihood therefore
disfavors all additive explanations \cite{Gordon05} (unless the explanation
helps significantly with some aspect of structure seen at higher $\ell$).

Explaining the observed correlations in terms of foregrounds is difficult.  
The combined quadrupole and octopole map suggests a foreground source which form a
plane perpendicular to the ecliptic.  It is clear neither how to form such a
plane, nor how it could have escaped detection by other means.  This planar
configuration means that single anomalous hot or cold spots do not provide an
adequate explanation for the observed effects.

The final possibility is that systematic effects remain in the analysis of the
WMAP data.  Two systematic corrections have already been introduced between the
one-year and the three-year data releases: the improved radiometer gain model,
which includes effects correlated with the ecliptic and with the seasons (and
hence the equinoxes); and the galaxy bias correction to the ILC.  We have
discussed the roles of both of these in reducing somewhat the statistical
significance of the orthogonality of the quadrupole and octopole planes to the
ecliptic.  But other systematic effects may remain.  For example, a suggestion
has been made \cite{Freeman:2005nx} that observed north-south power asymmetries
\cite{NS_asymmetry} could be caused by a small error (within the $1\sigma$
error bars) in the solar velocity with respect to the CMB. However, this would
not explain the correlations in full sky maps noted in this paper
\cite{Freemanprivate}.  Other possibilities have been put forward, for example
a correlation between the scanning pattern and the beam asymmetry
\cite{Meyerprivate}, but have not been carefully evaluated.

If indeed the observed $\ell=2$ and $3$ CMB fluctuations are not cosmological,
there are important consequences. Certainly, one must reconsider \cite{SSHC}
all CMB results that rely on low $\ell$s, such as the normalization, $A$, of
the primordial fluctuations and any constraint on the running $d n_s/d\log{k}$
of the spectral index of scalar perturbations (which, as noted in \cite{VSA},
depended in WMAP1 on the absence of low-$\ell$ TT power). Moreover, the
CMB-galaxy cross-correlation, which has been used to provide evidence for the
Integrated Sachs-Wolfe effect and hence the existence of dark energy, also gets
contributions from the lowest multipoles (though the main contribution comes
from slightly smaller scales, $\ell\sim 10$). Finally, note that, even though
we have discussed only alignments at $\ell=2$ and $3$, it is possible -- even
likely -- that the underlying physical mechanism does not cut off abruptly at
the octopole, but rather affects $\ell=4, 5, \ldots$.  Indeed, several pieces
of evidence, though less statistically convincing than those discussed here,
have been presented for anomalies at $l>3$.  If the values of $C_\ell$ for
$\ell>3$ are called into doubt, then so are the values of further cosmological
parameters, including the optical depth to the last scattering surface $\tau$,
the inferred redshift of reionization, and the rms mass fluctuation amplitude
$\sigma_8$.

Of even more fundamental long-term importance to cosmology, a non-cosmological
origin for the currently-observed low-$\ell$ microwave background fluctuations
is likely to imply further-reduced correlation at large angles in the true CMB.
As shown in Section 3, angular correlations are already suppressed compared to
$\Lambda$CDM at scales greater than 60 degrees at between $99.85$\% and
$99.97\%$C.L.\ (with the latter value being the one appropriate to the cut sky
ILC123).  This result is \textit{more} significant in the year 123 data than in
the year 1 data.  {\em The less correlation there is at large angles, the
poorer the agreement of the observations with the predictions of generic
inflation.  This implies, with increasing confidence, that either we must adopt
an even more contrived model of inflation, or seek other explanations for at
least some of our cosmological conundrums. } Moreover, any analysis of the
likelihood of the observed ``low-$\ell$ anomaly'' that relies only on the (low)
value of $C_2$ (especially the MLE-inferred) should be questioned.  According
to inflation $C_2$, $C_3$ and $C_4$ should be independent variables, but the
vanishing of ${\cal C}(\theta)$ at large angles suggests that the different
low-$\ell$ $C_\ell$ are not independent.

Another striking fact seen in Table~\ref{tab:Shalf} is that the quadrupole and
octopole of cut-sky maps, as well as $S_{1/2}$ statistics, are significantly
lower than in the full-sky ILC maps, both for year 1 and year 123.  This seems
to be because of a combination of effects --- the ``galactic bias correction''
and the time-dependent radiometer gain model certainly, but perhaps also just
the minimal variance procedure used to build the ILC.  
This reflects the fact that the synthesized full sky maps show correlations at
large angles that are simply not found in the underlying band maps.  This does
not seem reasonable to us --- that one starts with data that has very low
correlations at large angles, synthesizes that data, corrects for systematics
and foregrounds and then concludes that the underlying cosmological data is
much more correlated than the observations --- in other words that there is a
conspiracy of systematics and foreground to cancel the true cosmological
correlations.

This strongly suggests to us that there remain serious issues relating to the
failure of statistical isotropy that are permeating the map making, as well as
the extraction of low-$\ell$ $C_\ell$.  

At the moment it is difficult to construct a single coherent narrative of the
low $\ell$ microwave background observations.  What is clear is that, despite
the work that remains to be done understanding the origin of the observed
statistically anisotropic microwave fluctuations, there are problems looming at
large angles for standard inflationary cosmology.

\bigskip

The authors thank M. Dennis, O. Dor\'{e}, P. Ferreira, P. Freeman, C. Gordon,
G. Hinshaw, W. Hu, K. Inoue, S. Meyer, R. Nichol, H. Peiris, J. Silk, D. Spergel, and
R. Trotta for useful discussions.  The work of CJC and GDS has been supported
by the US DoE and NASA.  GDS has also been supported by the John Simon
Guggenheim Memorial Foundation and by a fellowship from the Beecroft Centre for
Particle Astrophysics and Cosmology.  DH is supported by the NSF Astronomy and
Astrophysics Postdoctoral Fellowship under Grant No.\ 0401066. GDS acknowledges
Maplesoft for the use of Maple.  We have benefited from using the publicly
available CAMB \cite{CAMB} and Healpix \cite{healpix} packages.  We acknowledge
the use of the Legacy Archive for Microwave Background Data Analysis
(LAMBDA). Support for LAMBDA is provided by the NASA Office of Space Science.


\end{document}